\documentclass{ar-1col}
\usepackage[numbers]{natbib}
\usepackage{amsmath}
\newcommand{\STRUT}{\rule{0in}{3.0ex}}
\jname{Xxxx. Xxx. Xxx. Xxx.}
\jvol{AA}
\jyear{YYYY}
\doi{10.1146/((please add article doi))}

\begin{document}

\markboth{A. C. Hayes and Petr Vogel}{Reactor Neutrino Spectra}

\title{Reactor Neutrino Spectra}

\author{Anna C. Hayes,$^1$ and Petr Vogel$^2$ 
\affil{$^1$T-2 Theoretical Division MS283, Los Alamos National Laboratory, Los Alamos, NM 8545, USA ; email: anna\_hayes@lanl.gov }
\affil{$^2$Kellogg Radiation Laboratory 106-38, California Institute of Technology, Pasadena, CA 91125, USA ; email: pvogel@caltech.edu}}

\begin{abstract}
We present a review of the antineutrino spectra emitted from reactors. 
Knowledge of these and their associated uncertainties
are crucial for  neutrino oscillation studies. 
The spectra used to-date have been determined by either conversion of
measured electron spectra to antineutrino spectra or 
by summing over all of the thousands of transitions that makeup the spectra using modern
databases as input.
The uncertainties in the subdominant corrections to beta-decay plague both methods, 
and we provide estimates of these uncertainties.
Improving on current knowledge of the antineutrino spectra from reactors will require new experiments.
Such experiments would also address the so-called reactor neutrino anomaly and the possible origin of the
shoulder observed in the antineutrino spectra measured in recent high-statistics reactor neutrino experiments.
\end{abstract}

\maketitle

\tableofcontents

\section{INTRODUCTION}
Nuclear reactors are intense, pure, and controllable sources of low energy electron antineutrinos.  
They have been frequently, and very successfully, used
in studies of fundamental neutrino properties. They will continue to play this role in the foreseeable
future. It is, therefore,
 important to understand the corresponding $\bar{\nu}_e$ flux, its energy distribution, and the associated
uncertainties in as much detail as possible. Here we review the work devoted to this issue.

The existence of neutrinos was suggested by Pauli already in 1930, in order to resolve the then apparent
energy and angular momentum non-conservation
in nuclear beta decay. Yet, the proof that neutrinos are real particles had to wait until 1953-1959, when
Reines and Cowan \cite{Reines,Reines2} 
detected the electron antineutrinos
emitted by a nuclear reactor. 
That fundamental experiment was the beginning of the field of neutrino exploration using 
reactor antineutrinos.       

The most important discovery in neutrino physics to-date is the existence of neutrino oscillations and 
by consequence the finite, albeit very small, rest mass of the neutrino. 
To explore oscillations with the early reactor experiments detectors were placed at distances $L \le 100$ m
\cite{ill,goesgen,rovno,krasnoyarsk1,Declais,bugey,bugey3,sc} and the observed $\bar{\nu}_e$ spectra were 
compared with that expected, the latter being based
on the then accepted evaluation. Neutrino oscillations, i.e. variation of the spectrum with the distance 
from the reactor, were not observed in these short baseline experiments, in agreement with 
our present knowledge of the three-neutrino oscillation phenomenology. Later reactor experiments 
\cite{chooz1,pv1,pv2} at larger distances ($\sim$ 1 km)
established an important upper limit for the mixing angle $\theta_{13}$, showing that this mixing angle 
is substantially smaller than the other two 
mixing angles, $\theta_{12}$ and $\theta_{23}$, the latter being
reasonably well determined at that time. Interpretation of results from these pioneering experiments 
was directly dependent on knowledge
of the reactor neutrino flux and spectrum. 

More recent, and still running, reactor neutrino experiments \cite{db1,db2,reno,dbch1,dbch2} are 
devoted primarily to the determination of the mixing
angle $\theta_{13}$ with the characteristic distance from the reactor of $\sim$ 1 km, 
i.e., near the corresponding oscillation minimum. 
In order to avoid, or substantially reduce, dependence on
detailed knowledge of the reactor spectrum, these experiments use two essentially identical detectors, with one or more 
placed relatively close to the reactors and the
other one (or several) further away. By comparing the signals at two distances it became possible 
to determine the oscillation signal corresponding
to the angle $\theta_{13}$ with very good accuracy. The detectors employed in these experiments are substantially larger than those 
in the previous generation of experiments and, thus, the statistical 
accuracy of the spectrum determination is substantially better. Although not the original 
intent, these modern experiments provide a detailed test of the
absolute reactor $\bar{\nu}_e$ flux and energy spectrum, and they raise new questions about our understanding of the expected spectra.


For precision reactor neutrino studies accurate knowledge of the reactor neutrino flux 
and spectrum is important. This issue became more pressing with
the reevaluation of the spectra in 2011 in Refs. \cite{mueller,huber}, which resulted in the upward 
revision of the expected reactor antineutrino signal by $\sim 6\%$. These revisions suggested
that all above mentioned experiments are missing approximately $6\%$ of the signal, 
independent of the distance from the reactor, beginning at $L \ge$ 10 m. 
This shortfall has become known as the ``reactor anomaly" and it has been interpreted 
\cite{mention} as a possible indication of the existence of an additional, fourth, necessarily sterile, light neutrino of mass $O$(1 eV), 
that becomes observable through subdominant mixing with the active neutrinos. 
If confirmed, this would be a discovery of fundamental importance. However, the sterile neutrino interpretation 
of the anomaly  hinges on the accuracy 
of the expected reactor neutrino flux.

\section{NUCLEAR REACTORS AS ELECTRON ANTINEUTRINO SOURCES}

Nuclear reactors derive their power from the fission of U and Pu isotopes and 
from the radioactive decay of the corresponding fission fragments. 
The beta decay of the fragments is the source of the electron antineutrinos.
The total antineutrino spectrum can be expressed as a sum over the spectra for the
dominant fissioning actinides,
\begin{equation}
S(E_{\nu}) = \Sigma_i f_i \left( \frac {dN_i}{dE_{\nu}} \right) ~,
\label{spectrum1}
\end{equation}
where $f_i$ is the number of fissions from actinide $i$ and $dN_i/dE_{\nu}$ is 
the cumulative $\bar{\nu}_e$ spectrum of $i$ normalized per fission.
Thus, as a first step, the parameters $f_i$ must be determined, which requires detailed information from the
reactor operator, including the total thermal power and the linear combination of actinides contributing to the power. 
The total 
reactor thermal energy $W_{th}$ and the parameters $f_i$ are related through
\begin{equation}
W_{th} = \Sigma_i f_i e_i ~,
\end{equation} 
where $e_i$ is the effective thermal energy per fission contributed by each actininde $i$. 
In power reactors 99.9\% of the power comes from the fission
of $^{235}$U, $^{239}$Pu, $^{241}$Pu and $^{238}$U, and only these isotopes 
are considered. The corresponding effective energies per fission are determined from the energy released in fission, minus the energy carried off by the antineutrinos, plus the energy produced by neutron  captures on the reactor materials. 
The evaluated \cite{james, kopeikin, ma} energies, 
$e_i$, are given in Table \ref{energyperfission}. The corresponding estimated uncertainties are 0.25-0.5 \%.
 
\begin{table}[h]
\caption{Transforming the thermal power into the fission rate  
(all energies in MeV/fission). 
Columns 2-4 are from Ref. \cite{james}.}
\label{energyperfission}
\begin{center}
\begin{tabular}{@{}|l|c|c|c|c|c|@{}}
\hline
Nucleus & Energy from &  Without $\nu$ &$E_{TOT}$ including & &  \\
  & mass excess &  & n-captures  & $E_{TOT}$ Ref. \cite{kopeikin}&$E_{TOT}$ in Ref.\cite{ma}\\
\hline
$^{235}$U & 202.7$\pm$0.1 & 192.9$\pm$0.5 & 201.7$\pm$0.6 & 201.92$\pm$0.46i&202.36$\pm$0.26\\
$^{238}$U  & 205.9$\pm$0.3 & 193.9$\pm$0.8 & 205.0$\pm$0.9 & 205.52$\pm$0.96&205.99$\pm$0.52\\
$^{239}$Pu & 207.2$\pm$0.3 & 198.5$\pm$0.8 & 210.0$\pm$0.9 & 209.99$\pm$0.60&211.12$\pm$0.34\\
$^{241}$Pu & 210.6$\pm$0.3 & 200.3$\pm$0.8 & 212.4$\pm$1.0 & 213.60$\pm$0.65&214.26$\pm$0.33\\
\hline
\end{tabular}
\end{center} 
\end{table}

The data for $W_{th}$ are usually available as function of time, while $f_i$, which are
typically expressed as the relative fractions
$f_i/F$, where $F$ is the total number of fissions, are obtained by (often proprietary) simulations. 
The neutrino spectrum in eq. (\ref{spectrum1})
can be expressed as
\begin{equation}
S(E_{\nu}) = \frac{W_{th}}{\Sigma_i (f_i/F) e_i} \Sigma_i \frac{f_i}{F} \left( \frac {dN_i}{dE_{\nu}} \right) ~.
\label{spectrum2}
\end{equation}
In writing eq.(\ref{spectrum2}) we are implicitly assuming that long-lived fission fragments not decaying in
 equilibrium have been corrected for. This issue is discussed in more detail below.
There can also be contributions to the antineutrinos emitted from the reactor complex from the radioactive spent fuel stored there. This correction,
which involves low-energy antineutrinos, is taken into account in oscillation experiments using inventory information supplied by the power company.

In order to determine the uncertainty in the $\bar{\nu}_e$ spectrum, it is necessary to determine 
the uncertainties in $W_{th}$ and in $f_i/F$, as well as their correlations.
The thermal power of a reactor is most accurately determined by temperature measurements in the coolant and
the calculation of water flow rates and the energy balance 
around the reactor vessel or steam generator.
It has been estimated, e.g. in Ref. \cite{djurcic} and in the references quoted there, that the total uncertainty can be 
as low as $\sim$ 0.5-0.7\%, although more typically values of the order of 2\% are quoted for $W_{th}$, 
and  government
regulations often allow a safety margin at this higher level.    

\begin{figure}[htb]
\begin{center}
\includegraphics[width=3.5in]{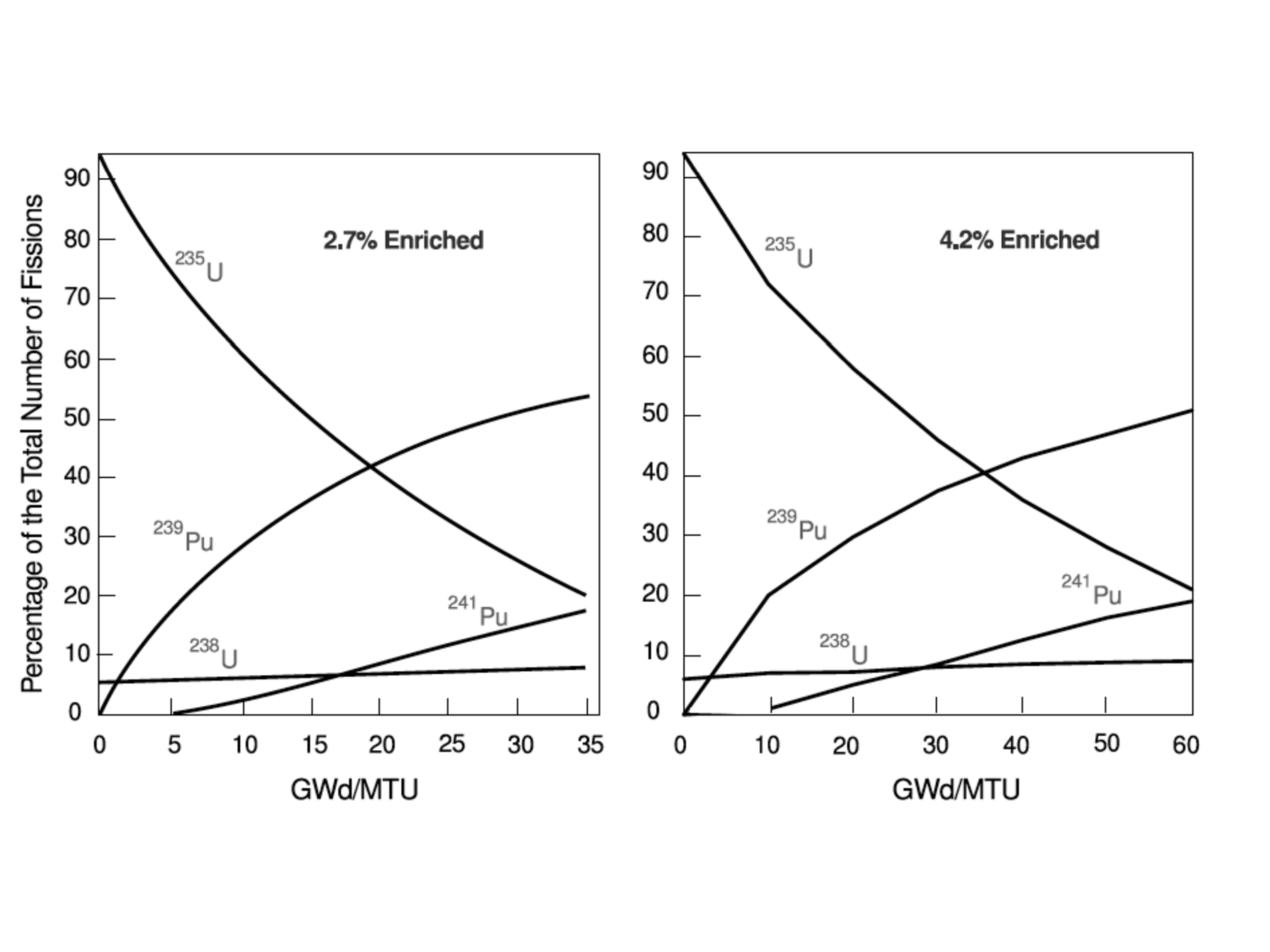}
\caption{
The evolution of the fuel composition for a pressurized water reactor over the reactor cycle, from Nieto {\it et al.} \cite{nieto}. 
The x-axis, GW-days per metric ton of in-going uranium fuel, is proportional to the number of fissions.
As the fuel enrichment increases, the burn curves to not change significantly, rather the scale on the 
x-axis becomes expanded. 
}
\label{fig:burnhist}
\end{center}
\end{figure}

\begin{figure}[ht]
\includegraphics[width=3.0in]{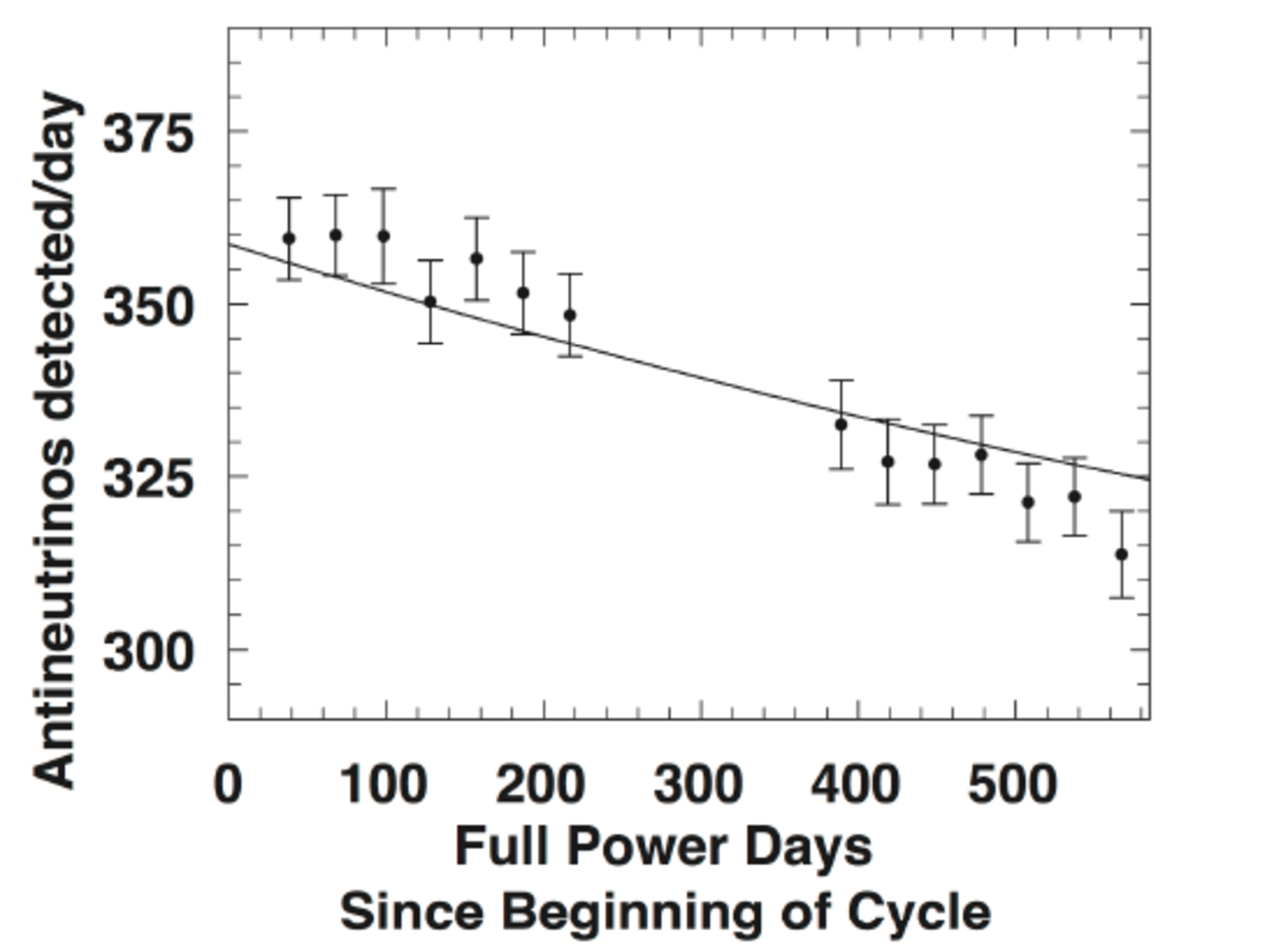}
\caption{
The change in the total number of antineutrnos emitted over the course of a reactor fuel cycle for a pressurized water reactor, resulting from the
in-growth of $^{239}$Pu, from Bowden {\it et al.} \cite{bowden}}
\label{fig:signaldrop}
\end{figure}

Fuel management, i.e, fuel recycling and the repositioning of fuel rods within the reactor core, is quite reactor design specific.  
For example, CANDU reactors involve frequent recyling of fuel  and the fuel composition {$f_i$} is kept close to constant.
In pressurized water reactors, on the other hand, during each reactor cycle, which typically lasts about a year,
the fuel composition is constantly changing; the $^{235}$U is being depleted, Pu is being bred, and 
the Pu fission fraction is increasing. Though it represents the vast majority of the fuel, $^{238}$U only contributes 
$~\sim10\%$ of  the total fission because it is a threshold 
fission actinide, and the percentage of fissions 
from $^{238}$U varies very slowly.  
The fraction of fissions from $^{238}$U depends on the enrichment of the fuel
and on the ratio of the thermal to fast neutron flux, which are two anti-correlated parameters in the reactor design. 
Fig. \ref{fig:burnhist} shows the variation in the fuel fission fraction $f_i/F$ as a function of burnup,
GWdays per metric ton of in-going uranium fuel, which is equivalent to number of fissions.
The left panel of this figure is for fresh 2.7\% $^{235}$U fuel. If the fuel enrichment is increased, the shape and magnitudes of the
curves do not
change signifcantly, rather the scale on the x-axis is expanded (right panel of Fig. \ref{fig:burnhist}), e.g., 
the burnup at which the fraction of fissions from $^{239}$Pu overtakes that from $^{235}$U is higher for higher enrichment.

In standard pressurized water reactors, 
at the end of each cycle about one-third of the fuel rods (those that have been burned for three cycles) are replaced with fresh fuel, 
and the position of many of the remaining partly 
burned rods is changed in order to keep the neutron flux across the reactor as close to flat as is possible. 
The average fuel 
composition as a function of time is simulated by detailed reactor burn codes, that often include a Monte Carlo treatment of the neutron transport.
The codes are normally specific to the reactor in question and checked and fine tuned by comparisons to spent fuel isotopics.
For this reason, operators can quote the
fractions $f_i/F$ to higher accuracy that would be possible by independent untuned simulations, where isotopics of major actinides in spent fuel are reproduced at the $\sim3\%$ level and that of fission fragments considerably less accurately.  
The magnitude of $\bar{\nu}_e$ spectra from the fisision 
of different actinides are different, the total contribution per fission from 
$^{235}$U is about 45\% higher than from $^{239}$Pu 
and about 60\% lower than $^{238}$U. Thus, the total antineutrino signal per fission 
can change during the reactor burn cycle. However, that variation is relatively small, and the 
uncertainty related to the uncertainties in $f_i/F$ is less important
than the uncertainty in the reactor thermal power $W_{th}$.

The typical variation in the antineutrino signal as a function of burnup for a pressurized water reactor is shown in Fig. \ref{fig:signaldrop}.
The data agree quite well with the prediction and the overall effect is an
$\sim$ 10\% decrease of the count rate during a fuel cycle of about 550 days. This decrease, caused by the 
changes of the fuel composition, has to be quantitatively accounted
for in oscillation experiments. 
Assuming that the reactor power or neutron flux is known independently, this change
can also be used for remote monitoring of the operational 
status of a  nuclear reactor. The issues determining the expected antineutrino spectra and their uncertainties for a declared burn history are key to the subfield
of so-called ``Applied Antineutrino Physics" 
\cite{monitoring}, and they are clearly intimately related to the issues of
this review. However, we will not discuss this application in any detail here.      

In fission each actinide nucleus is split into two, usually unequal mass, fragments. 
In the case of $^{235}$U, for example, the double hump mass fragment 
distribution peaks at A = 94 and 140, respectively. The stable nuclei with those masses are 
$^{94}$Zr and $^{140}$Ce that have 98 protons and 136 neutrons
together. The initial system has, however, 92 protons and 142 neutrons. To reach stability, 
therefore, six neutrons have to transformed into six protons. That
can be accomplished only by weak interaction $\beta$ decays, in which six 
electrons and six electron antineutrinos are emitted. This is a general
result for all reactor fuels; there are $\sim$ 6 $\bar{\nu}_e$ per second emitted per fission, so a typical 
reactor emits $\sim 6 \times 10^{20}$ electron antineutrinos per
each GW of the thermal energy power. The cascade of 
$\beta$ decays of the fission fragments is a consequence of the general increase 
of the neutron to proton ratio with increasing mass. The fission fragments,
with masses near half of the initial nuclear mass, are neutron rich and hence they $\beta$ decay, 
with a typical cascade of three decays each.

The $\beta$ decays, the source of the reactor $\bar{\nu}_e$, are  not instantaneous; 
they have finite lifetimes. As a consequence the
spectrum requires certain time interval from the beginning of the fission process to 
reach a steady equilibrium. The time needed to reach equilibrium is
different for $\bar{\nu}_e$ of different energies, typically being shorter for higher energies.

When using the
reactor neutrinos to study neutrino oscillation, the neutrino capture on protons, 
$\bar{\nu}_e + p \rightarrow e^+ + n$ is almost exclusively used
for neutrino detection. That reaction has a threshold, in the laboratory frame where the protons are at rest,
$E_{thr} = [( M_n + m_e )^2 - M_p^2]/ 2 M_p$  = 1.806 MeV.
Antineutrinos above this threshold mostly come from nuclei with relatively short half-lives 
that reach  equilibrium within a few hours.
  However, there are some exceptions; there are six 
fission fragments with sizable fission yield, 
and Q $>$ 1.8 MeV, $^{97}$Zr, $^{132}$I, $^{93}$Y, $^{106}$Ru, $^{144}$Ce, and $^{90}$Sr. The first three of 
them reach equilibrium within $\sim$ 10 days,
the next two have half-lives of 367 and 284 days, and $^{90}$Sr has $T_{1/2}$ = 28.8 years and decays into 
$^{90}$Y with Q= 2.28 MeV. The effects of nonequilibrium
is discussed in Ref. \cite{kopeikin2}. 

For $^{235}$U above $\sim$ 3 MeV of the neutrino energy equilibrium is
reached  within one day. However, at the detection threshold it
takes about 100 days to reach 1\% stability. When testing the spectra using shorter
irradiation times it is therefore necessary to correct for such off equilibrium effects.

\section{THEORETICAL DETERMINATION OF THE REACTOR $\bar{\nu}_e$ 
FLUX AND SPECTRUM}
\label{section-theory}

There are two complementary ways to determine the expected electron antineutrino 
spectrum of a nuclear reactor, the {\it `ab initio'} summation and the electron spectrum conversion 
methods. 

Assuming that the thermal power $W_{th}$, the normalized 
fission fractions $f_i/F$, and the energy per fission $e_i$ of each fissioning isotope $i$ 
are known or determined, the total $\bar{\nu}_e$ spectrum  in
eq. (\ref{spectrum2}) requires detailed knowledge of the individual fission spectra 
$dN_i/dE_{\nu}$ for each of the four fuels ($^{235}$U, $^{238}$U, $^{239}$Pu, 
and $^{241}$Pu). 
It is usually assumed that these individual spectra depend only on the nuclear 
properties of the  
fissioning isotopes and their fission fragments for thermal (0.025 eV) neutron fission in the case of
 $^{235}$U, $^{239}$Pu
and $^{241}$Pu, and fast fission for $^{238}$U. This might not be completely 
accurate, since the fission fragment yields, i.e. the distribution of the fission fragments, depends to some extent 
on the reactor dependent energy shape of the neutron flux. 
Keeping this caveat in mind, we next discuss how the spectra $dN_i/dE_{\nu}$ are determined.

\begin{figure}[htb]
\begin{center}
\includegraphics[width=3.0in]{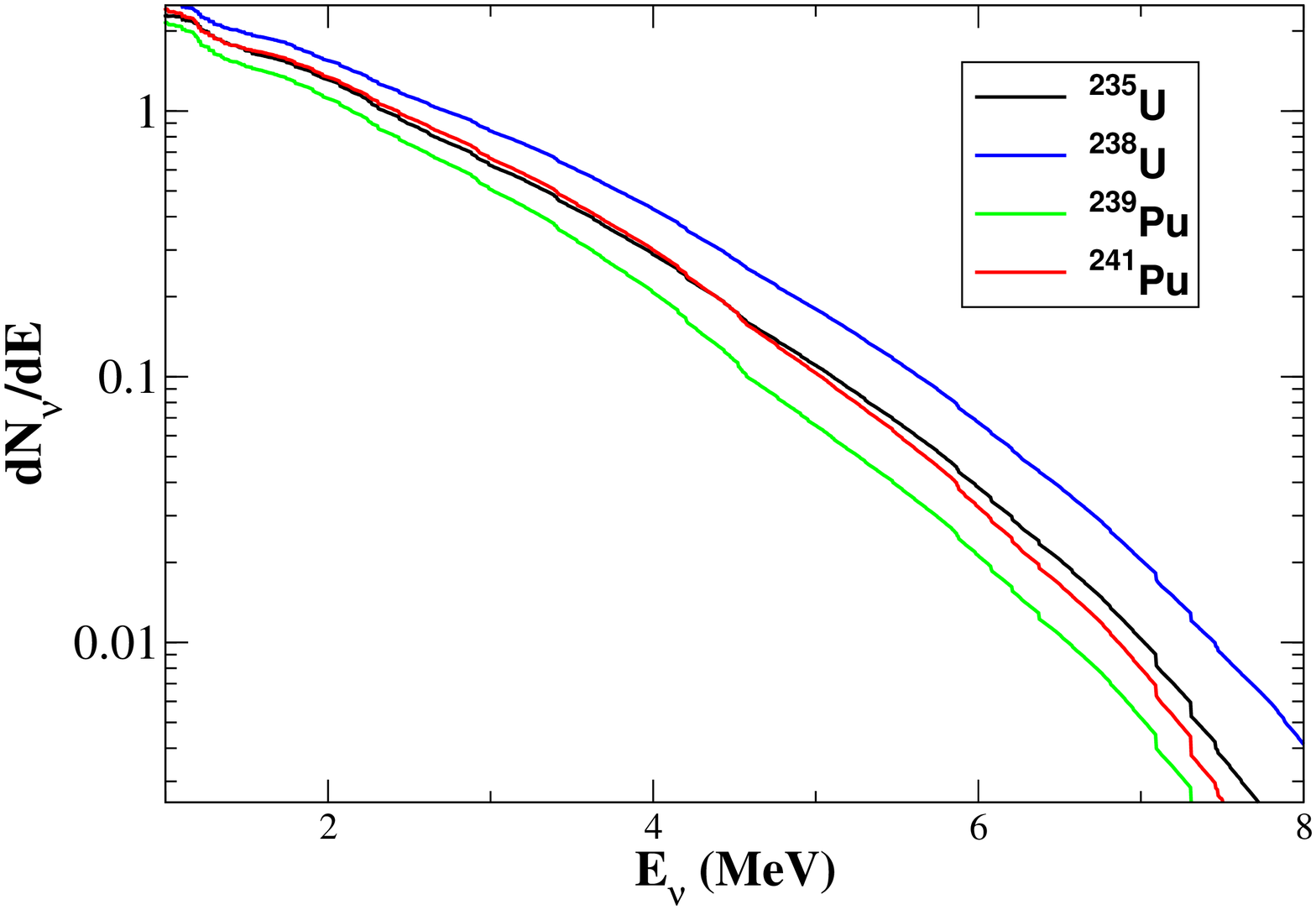}
\caption{
The antineutrino spectra for the four actinides determining the total antineutrino flux emitted from reactors. 
The fission yields were taken from JEFF-3.1.1 and the decay data, 
included the modeled data for unmeasured spectra, from ENDF/B-VII.1.} 
\label{fig:antispect}
\end{center}
\end{figure}

In the {\it `ab initio'} approach the aggregate fission antineutrino spectrum is determined by
summing the contributions of all $\beta$-decay branches of all fission
fragments
\begin{equation}
\frac{dN_i}{dE_{\bar{\nu}}} = \Sigma_n Y_n (Z,A,t) \Sigma_{n,i} b_{n,i}(E_0^i) 
P_{\bar{\nu}} (E_{\bar{\nu}}, E_0^i, Z) ~,
\label{fun1}
\end{equation}
where $Y_n (Z,A,t)$ is the number of $\beta$ decays of the fragment $Z,A$ at time $t$, 
and the label $n$ characterizes each fragment by whether it is in its ground state or an isomeric state.
After sufficient burn time the quantity $Y_n$ converges to the cumulative fission yield and is independent
of time.
Most fission fragments are produced by two mechanisms; first they are produced directly in the fission process with a
so-called independent yield, and second they are produced as the beta-decay daughter of a more 
neutron-rich fission fragment of the same mass number. 
The sum of the independent and beta-decay production of a fission fragment is its cumulative yield, and, once in equilibrium,
the cumulative yield determines the contribution of a given fragment to aggregate fission antineutrino spectrum.
The branching ratios $b_{n,i}(E_0^i)$ are characterized by the endpoint energies $E_0^i$.
They are normalized to unity, $\Sigma_i b_{n,i}(E_0^i) = 1$, unless the fragment decays by an additional mode other than beta decay. 
Finally, the function $P_{\bar{\nu}} (E_{\bar{\nu}}, E_0, Z)$ is the normalized 
$\bar{\nu}_e$ spectrum shape for the branch $n,i$. An analogous formula holds for the corresponding aggregate fission
electron spectrum, where $E_{\bar{\nu}}$ in the individual spectra $P$ must be replaced by
$E_e = E_0^i - E_{\bar{\nu}}$, since the nuclear recoil can be neglected within the accuracy considered here.  
Fig. \ref{fig:antispect} shows the antineutrino spectrum predicted by the summation method,
using the JEFF-3.1.1 \cite{jeff} database fission fragment yields and the ENDF/B-VII.1 \cite{endf} decay library. The ENDF/B-VII.1 
decay library used here is that up-dated in ref. \cite{sonzogni} to improve important issues with the older database pointed out in \cite{fallot}.

In applying the summation technique and eq. (\ref{fun1}) several sources of uncertainty arise.
The fission yields $Y_n$ have been evaluated by several international database groups, but for many important fragments the yields
involve large  uncertainties. The branching ratios $b_{n,i}$ are also not known for all fragments, and nor are the
 quantum numbers (spins and parity) of all of the initial and final states.
The shape of the $\beta$ decay spectrum $P$ is well known for  allowed transitions ($\Delta I \le 1, 
\pi_i \pi_f = 1)$ transitions. For neutron rich fission fragments all of the allowed transitions are Gamow-Teller transitions, determined by the operator $\sigma\tau$.
However, $\sim30\%$ of the transitions making up the aggregate spectra are known to be so-called first forbidden transitions, $(\Delta I \le 2, \pi_i \pi_f = -1)$,
and involve nuclear structure dependent combinations of several more complicated operators.
In the cases of some first forbidden operators, the spectra involve shapes that are noticeably different from those for allowed transitions, 
as described in the subsection \ref{subsecs:forbidden}.
Finally, there are important, albeit small,
corrections to the beta-decay spectra arising from radiative, nuclear finite size, and weak magnetism effects, and these can also depend on the details of the transition,
as described in the subsection.
\ref{subsecs:allowed}.  
The difficulties of the {\it `ab initio'} method, and the corresponding uncertainties are described in the next section and
in the section on uncertainties.

The second method of determining the spectra $dN_i/dE_{\nu}$  begins with the experimentally measured
aggregate {\it electron} spectrum associated with the fission of each individual actinide $i$. 
The electron spectrum for thermal neutron fission of $^{235}$U, $^{239}$Pu and $^{24 1}$Pu were measured at ILL, Grenoble, France
in 1980's \cite{Schr2,Schr3,Schr4}. The results were republished with a finer grid of electron
energies recently in the Ref \cite{Haag1}. $^{238}$U fissions only with fast neutrons; its electron spectrum was
measured much later at the neutron source FRMII in Garching, Germany \cite{Haag}.
These experimentally determined electron spectra are automatically summed over all 
fission fragments and the corresponding 
$\beta$-decay branches, so no information on the fission yields and branching ratios is needed. 
It is necessary, however, to convert them into the $\bar{\nu}_e$ spectra. 
 It is also necessary to make the relatively small correction for the fact that the electron spectra were determined before
 full equilibrium was reached.

\begin{figure}[htb]
\begin{center}
\includegraphics[width=3.0in]{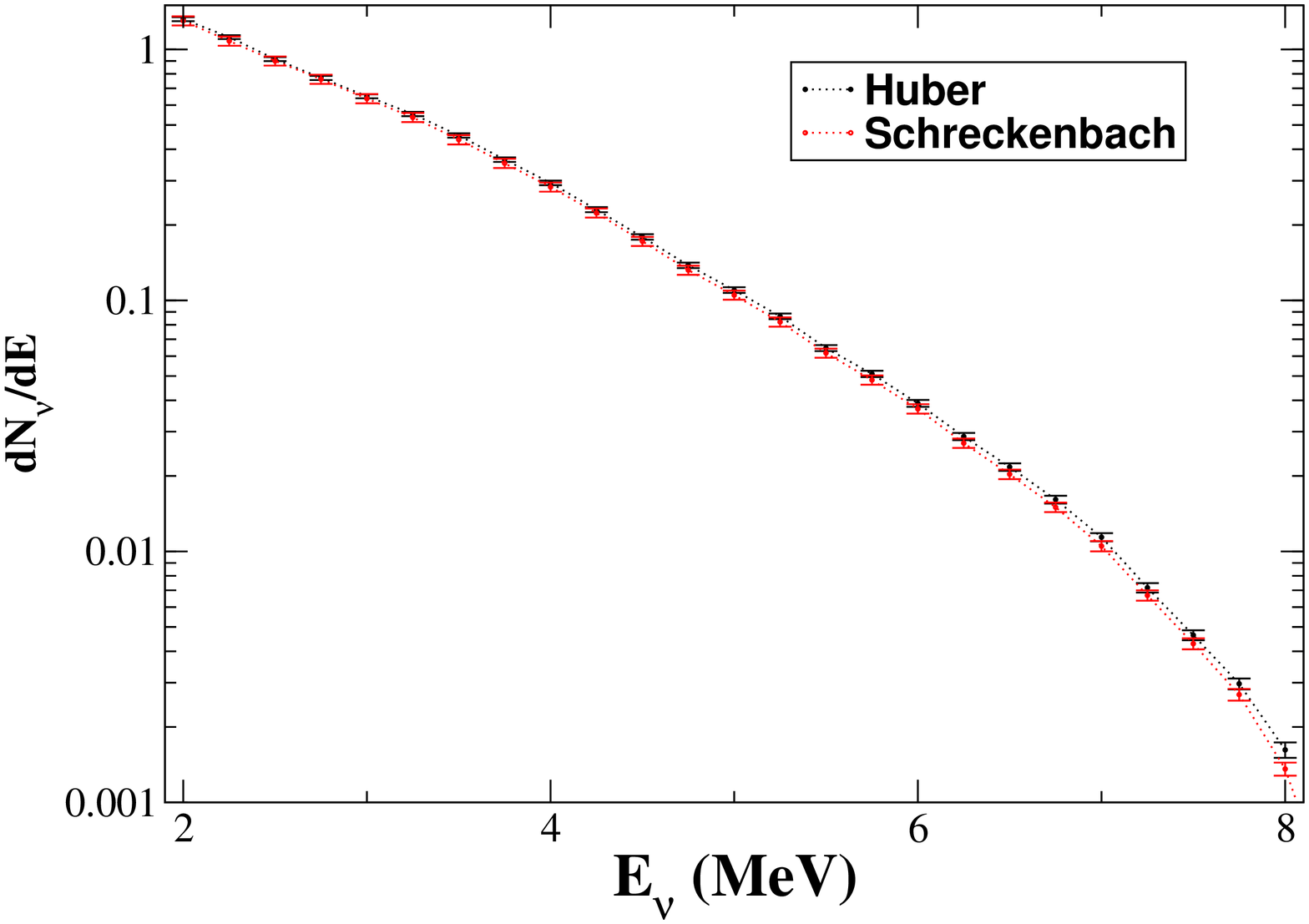}
\caption{
The antineutrino spectrum for $^{235}$U derived by converting \cite{huber,Schr2} the measured \cite{Schr2} electron spectrum.
The difference in the two derived  spectra arises from differences in the assumptions made about the subdominant
corrections to beta-decay. 
The uncertainty in the theoretical form of these corrections, discussed below and summarized in Table \ref {uncert}, 
are sufficently large that direct experimental measurements will be necessary to determine the correct normalization of
the antineutrino spectra to this accuracy.}
\label{fig:conv}
\end{center}
\end{figure}

To convert a measured aggregate electron spectrum into an antineutrino spectrum, the spectrum is binned over an  energy grid, 
with the grid defining a set of virtual end-point energies $E_0^i$. 
The total aggregate spectrum is then fitted in terms of the amplitudes $a_i$ for each virtual end-point energies,
$dN_i/dE_{e}=\Sigma_i a_i P(E,E_0^i, Z)$.
In principle, the position of the virtual end-point energies can also be part of the fit.
Thus, the aggregate electron spectra, which have been  measured in the energy window ($\sim2-8.5$ MeV),
are described by a sum of virtual $\beta$-decay branches of assumed spectral shapes. 
The conversion to the antineutrino
spectrum is then simply accomplished by replacing the energy $E_e$ in each branch by $E_0 - E_{\bar{\nu}}$.
The procedure guaranties that the experimental electron spectrum is well reproduced. It is also 
straightforward to test whether the convergence on the number of energy intervals is achieved, which typically requires less than 
30 intervals. 
However,
the converted $\bar{\nu}_e$ depends to some degree on the assumptions made about the 
spectrum shapes $P_i$, whether they correspond to allowed or forbidden transitions, their $Z$ dependence, and the form of the
corrections arising from nuclear finite size and weak magnetism. An example of the uncertainty in the converted spectrum is shown in Fig. \ref{fig:conv}. There can also be some dependence on the endpoint energies $E_0^i$.  
To avoid
sizable systematic errors when converting the electron spectrum it is necessary to use the
data bases and evaluate the dependence of the average nuclear charge Z on the endpoint
energy discussed in \cite{huber,vogel07}.

A hybrid combination of these two methods has been also used \cite{mueller}, in which equation (\ref{fun1})
is used for the fission fragments and $\beta$-decay branches where experimental data are available. 
Both electron and $\bar{\nu}_e$ spectra for this large subset of fission fragments are then evaluated. 
The difference between the measured electron spectrum \cite{Schr2,Schr3,Schr4} and the
evaluated partial electron spectrum is then converted into the $\bar{\nu}_e$ spectrum by the fitting procedure.
This hybrid method has the advantage of taking account of the measured properties of a large subset of the fission fragments, and using
experimental data to determine the energy dependence of forbidden transitions and the $Z$ dependence of $P_i$. 

In any of the methods, a necessary condition is a good understanding of the shape factors
$P_{\bar{\nu}} (E_{\bar{\nu}}, E_0^i, Z)$ of the individual $\beta$ decays, including
nuclear charge $Z$ and the end-points $E_0^i$, as well as the role of the allowed versus forbidden transitions.

\subsection{Corrections to the $\beta$ decay electron and $\bar{\nu}_e$ spectrum for the allowed $\beta$ transitions}
\label{subsecs:allowed}

The $\beta$-decay spectrum shape can be expressed as
\begin{equation}
P_{\bar{\nu}} (E_{\bar{\nu}}, E_0^i, Z) =
K p_e E_e (E_0 - E_e)^2 F(Z,E_e) C(Z,E_e) (1 + \delta(Z,A,E_e)) ~,
\end{equation}
where $K$ is the normalization factor (the function $P$ must be normalized to unity when integrated over $E_e$ and used in eq. (5)).
$p_e E_e (E_0 - E_e)^2$ is the phase space factor, $F(Z, E_e)$ is the Fermi function that takes into account the
effect of the Coulomb field of the daughter nucleus on the outgoing electron, and the shape factor $C(Z,E_e)$ accounts for the energy
or momentum dependence of the nuclear matrix elements. For the allowed decays, $C(Z,E_e) = 1$. Finally, the
function $\delta(Z,A,E_e)$ describes the subdominant corrections to the spectrum shape, to be discussed in 
detail below.

\subsubsection{ The Fermi Function}
  The Fermi function $F(Z,E_e)$ replaces the plane wave solution for the out-going electron with a Coulomb wave. 
It is straightforward to calculate $F$ under the assumption of a point nuclear distribution, which leads to a Fermi 
 function of the form,
\begin{equation}
F_0 (Z,E_e)  =  4 (2 p_e R)^{-2(1 - \gamma)} \left[ \frac{\Gamma(\gamma + i y)}{\Gamma(2 \gamma + 1)} \right]^2
e^{\pi y} ~,
\label{point}
\end{equation}
where $ \gamma = \sqrt{1 - (\alpha Z)^2}$ and $y = \alpha Z E_e/p_e$.
Here $R$ is the cut-off radius, normally taken to be the radius of the nucleus in units of the electron Compton wavelength.
The point Fermi function leads to a logarithmic divergence at 
$R=r = 0$ in eq.(\ref{point}). 
Of course, for a nucleus with a finite charge radius the solution to the Dirac equation for the wave function of the out-going electron  is finite everywhere. 

\subsubsection{The Finite Size Corrections}
It is not possible to derive a general and exact correct for the finite size correction to the  Fermi function.  
For this reason, different approximations have been made in the literature. These involve  assumptions about 
the nuclear charge $\rho_{CH}(r)$ and weak isovector transition $\rho_W(r)$ densities,
and perturbative expansions in $\alpha Z\left(\frac{ER}{\hbar c}\right)$ and/or in $q^2$ .

Holstein \cite{holstein} derived an analytic expression using a first order expansion  $\alpha Z\left(\frac{ER}{\hbar c}\right)$.
The result depends on {\it both} the
charge and weak densities. If the weak and charge densities are assumed to be the same $\rho_W=\rho_{CH}$ the finite size correction can be expressed \cite{friar1} in terms of the first Zemach moment \cite{zemach} $\langle r \rangle_{(2)}$, and is given
 in ref.\cite{hayes} as,
\begin{equation}
\delta_{FS} = -\frac{3}{2}\frac{Z\alpha}{\hbar c}\langle r \rangle_{(2)}\left ( E_e-\frac{E_\nu}{27}+\frac{m_e^2c^4}{3E_e}\right ) 
\label{holFS}
\end{equation}
The Zemach moment, 
\begin{equation}
\langle r \rangle_{(2)}=\int d^3r\rho_W(r)\int d^3r\rho_{ch}(s)|\vec{r}-\vec{s}|,
\label{zemach}
\end{equation}
is the first moment of the convoluted nuclear weak isovector transition density  and electromagnetic ground state charge densities.
Though the expression in  eq. (\ref{holFS})  is {\it exact} to order $\alpha Z$,  some assumption must be made about $\rho_W$ and $\rho_{CH}$
in calculating $\langle r \rangle_{(2)}$.

 Behrens {\it et al.} \cite{behrens} solved the finite size problem
numerically, including higher order terms in $\alpha Z$, but expanding the weak density $\rho_W$  to first order in $q^2$.
The evaluation proceeds in two steps. First, the singularity at the origin in $F_0(Z,E_e)$ is removed by replacing
it by the function $F(Z,E_e)$ based on the numerical solution to the Dirac equation for the outgoing electron
in a finite size Coulomb potential, and evaluating it at $r=0$
\begin{equation}
F(Z,E_e)  =  F_0 (Z,E_e) \cdot L_0(Z,E_e)  ~.
\end{equation}
The functions $L_0$, as well as $F(Z,E_e)$, are tabulated in \cite{behrens}.
A less accurate, but much simpler analytic form of $F_0 L_0$, accurate to about 1\% for $30 \le Z \le 70$ and $E_e \le$ 8 MeV,
is available in \cite{schenter}. 
In the second step, in addition to using $F(Z,E_e)$ in place of $F_0(Z,E_e)$, the finite nuclear size correction needs to be added.


\begin{figure}[htb]
\begin{center}
\includegraphics[width=3.0in]{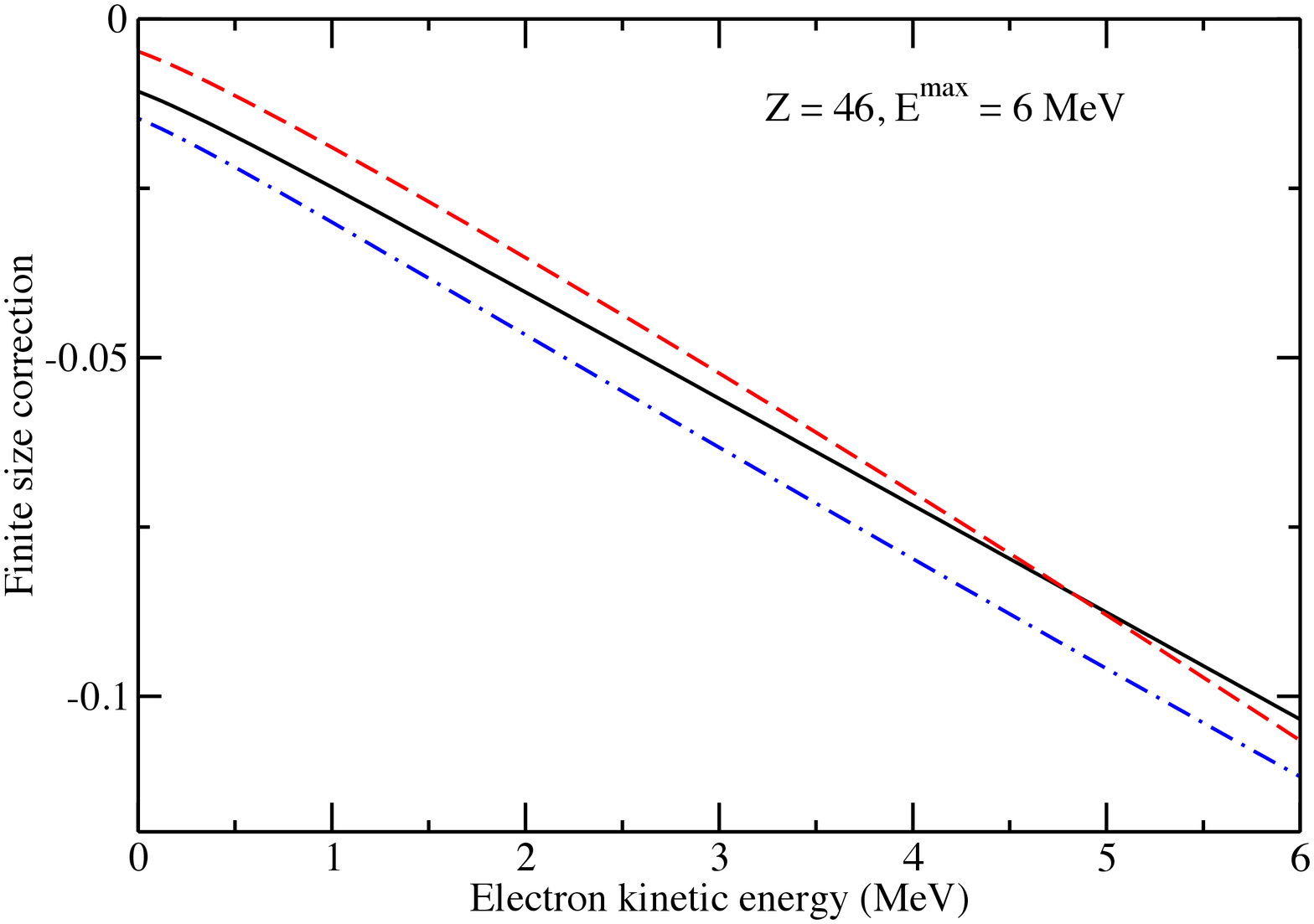}
\caption{The finite size correction $\delta_{FS}$ for $Z$ = 46 and maximum electron or $\bar{\nu}_e$ energy 6 MeV 
is plotted
versus the electron kinetic energy.
The full line is based on \cite{hayes}, the dashed line on \cite{wilkinson} and the dot-dashed one on \cite{vogel1}.
}
\label{Fig:finsize}
\end{center}
\end{figure}

The two expansion methods, in $\alpha Z$ versus $q^2$, lead to numerically similar results for uniform charge and weak density distributions of radius $R$, for which $\langle r \rangle_{(2)}= \frac{36}{35}R$, \cite{friar1}. 
In Ref. \cite{hayes}, uniform distributions were assumed and, ignoring constant terms independent of $E_e$, $\delta_{FS}$ becomes,
\begin{equation}
\delta_{FS}^{(1)} = - \frac{8}{5} \frac{Z \alpha R E_e}{\hbar c} \left( 1 + \frac{9}{28} \frac{m_e^2c^4}{E_e^2} \right)
\label{eq:finsize}
\end{equation} 
An identical value for $\delta_{FS}$ was obtained for allowed transitions
 in ref. \cite{holstein2}. In the latter reference it is noted that the magnitude of the 
finite size correction for uniform charge and weak densities is a factor
of 1.3 smaller than that obtained assuming surface densities.

Terms that are higher order in $\alpha Z$
introduce small corrections that scale with $R^2$, $R/E_e$ and matrix elements of the operators $r^2[Y_2 \times \vec{\sigma}]$ and $\vec{\sigma} \times \vec{l}$.
In ref. \cite{wilkinson} the approximation of Behrens {\it et al.} 
is expressed in terms of an empirical  analytic expression for allowed
Gamow-Teller transitions. 
That formula was also applied in the reactor spectrum evaluation in \cite{huber}.
A somewhat different but close formula was obtained in \cite{vogel1}, where an average for a uniform or surface weak density distribution was estimated in terms of the matrix element ratio $\frac{\langle \sigma r^2 \rangle}{\langle \sigma\rangle R^2}$, and the finite size correction written as, 
\begin{equation}
\delta_{FS}^{(2)} = -\frac{9}{10} \frac{Z \alpha R E_e}{\hbar c} \frac{\langle \sigma r^2 \rangle}{ \langle \sigma \rangle R^2} .
\end{equation}
For a uniform (surface) distribution $\frac{\langle \sigma r^2 \rangle}{ \langle \sigma \rangle R^2}=3/5 ~(1)$.
Ref. \cite{mueller} followed this form and used $\delta_{FS}^{(2)} = -\frac{9}{10} \frac{Z \alpha R E_e}{\hbar c}$.

In Fig. \ref{Fig:finsize} three of the different forms for 
the finite size corrections $\delta_{FS}$ that have been used in the literature for a  uniform density are compared. 
Only the energy dependent part of $\delta_{FS}$ is plotted; the energy independent component is irrelevant for the normalized
spectra. These differences, as well as the assumptions that must be made in evaluating  $\langle r \rangle_{(2)}$, suggest that a large uncertainty  needs to be assigned to $\delta_{FS}$ for allowed GT transitions.

\subsubsection{The Radiative Corrections}
The QED corrections of the first order in $\alpha$ to both the electron and $\bar{\nu}_e$ spectra 
in the $\beta$ decay have been evaluated in Refs. \cite{sirlin1, sirlin2}.
An earlier version can be found in \cite{batkin}. Only the energy dependent corrections to the  
electron and $\bar{\nu}_e$ spectrum are relevant;  only they affect the spectrum shape. 
They are of the form
\begin{equation}
\delta_{QED}^{\bar{\nu}} = \frac{\alpha}{2 \pi} h(E_e, E_0)~, ~~ \delta_{QED}^e =  \frac{\alpha}{2 \pi} g(E_e, E_0)~,
\label{eq:qed}
\end{equation}
where $E_{\bar{\nu}} = E_0 - E_e$ and the functions $h(E_e, E_0)$ and $g(E_e, E_0)$ are defined in \cite{sirlin1,sirlin2}.
We note that for the conversion of the electron spectrum to the $\bar{\nu}_e$ spectrum only the difference 
$h(E_e, E_0) - g(E_e, E_0)$ is  relevant.

\subsubsection{The Weak Magnetism Correction}
The interaction of the out-going electron with the magnetic moment of the daughter nucleus leads to a 
weak magnetism correction. 
The form of the correction is determined by the interference of the
magnetic moment distribution of the vector current $\vec{J}_V=\vec{\nabla}\times\vec{\mu}$
with the spin distribution $\vec{\Sigma}$ of the axial current. Thus, there is {\it no} weak magnetism correction to Fermi or 
pseudo-scalar ($0^-$) transitions.
 In
the non-relativistic approximation the correction depends on nuclear matrix elements of the operators $\vec{\sigma}$ 
and $\vec{l}$ and for GT transitions has the form \cite{holstein}
\begin{equation}
\delta_{WM} = \frac{4 E_e}{3 g_A M} \left(1 - \frac{m_e^2c^4}{2 E_e^2} \right)
\left[ \frac{ \langle \vec{l} \rangle}{ \langle \vec{\sigma} \rangle}
+ (\mu_p - \mu_n) \right] ~,
\end{equation} 
where $\mu_p - \mu_n = 4.7$ is the nucleon isovector magnetic moment. 
In principle, the matrix element ratio $\frac{ \langle \vec{l} \rangle}{ \langle \vec{\sigma} \rangle}$ needs to be evaluated separately for each transition. As an approximation,
one can use the truncated orbital current \cite{hayes}
\begin{equation}
\delta_{WM} \approx \frac{4}{3} \frac{\mu_p - \mu_n - 1/2}{g_A M} E_e  \left( 1 - \frac{m_e^2c^4}{2 E_e^2}
\right) ~\approx~ 0.5\%  E_e/{\rm MeV} ~,
\label{eq:wm}
\end{equation}
An analogous weak magnetism correction, without the relatively small term $m_e^2/2E_e^2$, was suggested in 
\cite{vogel1} and used in \cite{mueller,huber}. 
In light nuclei it is possible to test the leading order term of weak magnetism correction $\delta_{WM}$ through 
its relation to the
decay width of the $M1$ $\gamma$-ray transition for isobaric analog states. 
A list of these cases can be found in Ref. \cite{huber},
resulting in an average slope of $0.67 \pm 0.26$\%, in a fair agreement with the above formula. It is impossible, however,
to test $\delta_{WM}$ for the transitions of real interest, i.e. for the $\beta$ decay of fission fragments. The estimate above, 
therefore, must be assigned a sizable uncertainty.

\begin{figure}[htb]
\begin{center}
\includegraphics[width=2.5in]{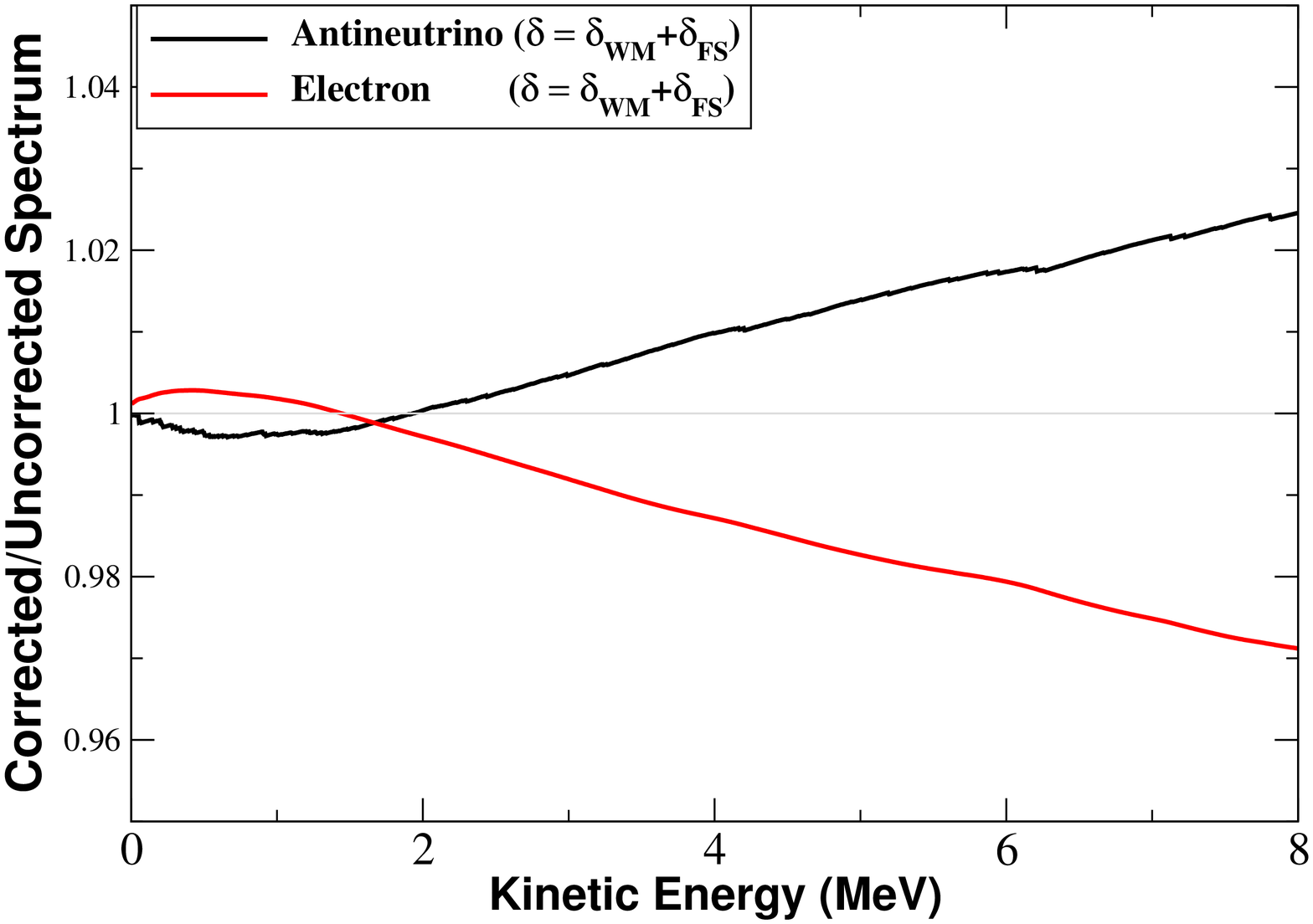}
\caption{The finite size $\delta_{FS}$ and weak magnetism $\delta_{WM}$ corrections result in an approximately
 linear increase (decrease) in the antineutrino (electron) spectrum above half the average end-point energy $\overline{E}_0/2$.
The figure shows the ratio of the spectra with and without these two corrections, using the summation method and the ENDF/B-VII.1 database. 
The form of the corrections used here are those given in eqs. (\ref{holFS}) and (\ref{eq:wm}), and the $Z$ and $A$ values involved 
are taken from the database.
}
\label{correct}
\end{center}
\end{figure}

The effect of the corrections on the electron and antineutrino spectra is summarized in Fig. \ref{correct}.
Since the spectrum for each fission fragment  must be normalized to unity when integrated over all energies, 
the corrections increase the aggregate spectrum at some energies and lower it at other energies.
In particular, below half the average end-point energy for all fission fragments, $\overline{E}_0/2$, 
the electron (antineutrino) spectrum is increased (decreased). 
Above $\overline{E}_0/2$, the electron (antineutrino) spectrum is decreased (increased). The approximate linear  form of $\delta_{FS}$ and $\delta_{WM}$ in energy causes the decrease (increase) 
above $\overline{E}_0/2$ to also be approximately linear and to have a slope $\mp\frac{1}{2}(\delta_{FS}+\delta_{WM})$.
A change in $\delta_{FS}$ or $\delta_{WM}$ to account for the uncertainties in these corrections
  would be directly 
reflected in a change in this
slope. This point is important in assessing the statistical significance of the reactor anomaly, to be discussed later. 
\ 

\subsection{First forbidden $\beta$ decays}
\label{subsecs:forbidden}

In the ground states of fission fragments the least bound protons and
the least bound neutrons are often
in states of opposite parity belonging to different oscillator shells. For this reason about
30\% of all $\beta$ decays contributing to the reactor neutrino spectrum are forbidden decays.
The forbidden decays tend to be more prevalent in the higher energy part of the aggregate spectra,
where the phase space advantage wins out over the suppression due to the forbiddenness of the transitions, 
the latter nominally scaling with $(pR)^2 \ll 1$.  

The selection rules for the first forbidden $\beta$ decays are $\pi_i \pi_f = -1$ and
$\Delta J \le 2$. 
In the leading order there are six relevant operators \cite{hayes}, which can be reduced in the number of independent operators by  invoking the conserved vector current and relating
operators proportional to $\vec{\nabla}/M\tau$  to the operator $\vec{r}\tau$.  
Two of the operators, $\gamma_5$
and $\vec{\alpha}$ involve emission
of the $s_{1/2}$ electrons, and hence the corresponding shape factor $C(Z,E) = 1$,
as is the case for allowed decay. However, the four additional operators involve the emission
of the $p_{1/2}$ state electrons, and  $C(Z,E) \ne 1$ in these cases. 
The shape factors $C(Z,E)$ for the six first forbidden operators are listed in Table \ref{six}.

\begin{table*}
\caption{The shape factors and leading-order weak magnetism corrections to allowed and first-forbidden decays
from \cite{hayes}. The top panel is for Gamow-Teller transitions. The shape factors for allowed and first-forbidden
Fermi beta decays are shown in the bottom panel. 
Nuclear operators $\vec{J}_V$ and $\rho_A$ are proportional to a nucleon
velocity $(p/M_N)$. CVC has been involked to replace them by the analogous
operators proportional to $E_0r$ for $\vec{J}_V$ , and a similar approximation has been made for the $\rho_A$ operators
proporional to $(p/M_N)$.
The weak magnetism correction for $\vec{J}_V$ involves the unknown overlap of very different $1^-$ matrix
elements and is therefore not listed. The nucleon isovector magnetic moment is $\mu_v =4.7$, $M_N$
is the nucleon mass, $g_A$ is the axial vector coupling constant, and $\beta = p_e/E_e$.
}
\vspace{4pt}
\begin{tabular}{lcccc}
Classification & $\;\Delta J^\pi\;$ & \; Oper. \; & Shape Factor $C(E_e)$ & \; Fractional Weak Magnetism Correction $\delta_{\mathrm{WM}}(E_e)$\\\hline\hline
Gamow-Teller:&&&&\\
Allowed &$1^+$&$\Sigma\equiv\sigma\tau$&1&$\frac{2}{3}\left[\frac{\mu_v-1/2}{M_Ng_A}\right](E_e\beta^2-E_\nu)$\\
 1$^{st}$ F.&$0^-$&$\left[\Sigma,r\right]^{0-}$&$p_e^2+E_\nu^2+2\beta^2E_\nu E_e$&0\\
\!\! 1$^{st}$ F. $\rho_A$ &$0^-$&$\left[\Sigma,r\right]^{0-}$&$\lambda\,  E_0^2$&0\\
 1$^{st}$ F. &$1^-$&$\left[\Sigma,r\right]^{1-}$&$p_e^2+E_\nu^2-\frac{4}{3}\beta^2E_\nu E_e$&$\;\;\;\;\left[\frac{\mu_v-1/2}{M_Ng_A}\right]\left[\frac{(p_e^2+E_\nu^2)(\beta^2 E_e-E\nu)+2 \beta^2 E_e E_\nu(E_\nu-E_e)/3}{(p_e^2+E_\nu^2-4\beta^2E_\nu E_e/3)}\right]$\\
Uniq. 1$^{st}$ F. &$2^-$&$\left[\Sigma,r\right]^{2-}$&$p_e^2+E_\nu^2$&\,\,$\frac{3}{5}\left[\frac{\mu_v-1/2}{M_Ng_A}\right]\left[\frac{(p_e^2+E_\nu^2)(\beta^2 E_e-E\nu)+2 \beta^2 E_e E_\nu(E_\nu-E_e)/3}{(p_e^2+E_\nu^2)}\right]$\STRUT \\[1.5ex] \hline 
Fermi:&&&&\\
Allowed  &$0^+$&$\tau$&1&0\\
\!\!\! 1$^{st}$ F. &$1^-$&$r\tau$& $p_e^2+E_\nu^2+\frac{2}{3}\beta^2E_\nu E_e$& 0\\[0.7ex]
 1$^{st}$ F. $\vec{J}_V$ &$1^-$&$r\tau$& $E_0^2$& -\\[0.7ex]
\hline
\label{six}
\end{tabular}
\end{table*}

In the case of $\Delta J = 2$, the unique first forbidden transition, only one operator can contribute,
and corresponding shape factor is $C(Z,E) =  p_e^2 + p_{\nu}^2$. For $\Delta J= 0$ two
operators contribute and for $\Delta J = 1$ three. 
In general, the overall shapes factor $C(Z,E)$ of such transitions
depends on the magnitude and sign of the matrix elements of the different forbidden operators contributing
to the transition.
The situation is often even more complicated since $|J_i - J_f| \le \Delta J \le J_i + J_f$ so that
in typical decays of an odd-A or odd-odd nucleus more than one $\Delta J$ contributes.
For the even-even nuclei $J_i  = 0$ and only one $\Delta J = J_f$ is relevant in that case

First forbidden $\beta$ decays often exhibit spectra of similar shape to
allowed decays. As pointed in ref. \cite{weidenmuller}, 
this is likely the case whenever the Coulomb energy of the emitted electrons is much larger
than its total energy at the nuclear radius, $\alpha Z/R \gg E_0/m_e$, with $R$ is expressed in the electron Compton
wavelength units. This limit is often referred to as the  $\xi$ approximation, \cite{weidenmuller}. 
However, taking as an example the important
decay of $^{92}$Rb with $\alpha Z/R = 19.2$, and $E_0 = 16.8 m_e$,  $\alpha Z/R \sim E_0$.
For this and many of the high $Q$-value decays that dominate the aggregate spectra above 5 MeV, the $\xi$ approximation
cannot be used as guidance. Nevertheless,
in the case  of $^{92}$Rb, at least, the measured $\beta$ spectrum \cite{rudstam}, dominated by the
$0^- \rightarrow 0^+$ ground state branch, has essentially an allowed shape. 

The QED or radiative corrections to the spectrum, $\delta_{QED}$, depend only on the emitted electron
energy. Hence $\delta_{QED}$, defined in the eq. (\ref{eq:qed}),
 is the same for forbidden transitions as it is for the allowed decays.

On the other hand, the weak magnetism corrections are operator dependent. They are listed for first forbidden transitions
in ref. \cite{hayes}.
As noted above, $\delta_{WM}$ vanishes for $\Delta J = 0$
operators. In particular, there is no weak magnetism correction for $0^- \rightarrow 0^+$ transitions, and such transitions
represent an important component of the antineutrino spectra, especially at high energy.
The weak magnetism correction also vanishes for the
vector current (Fermi) operator $\vec{r}$, which is one of the operators responsible for the $\Delta J = 1^-$
transitions. 
Thus, in the absence of detailed calculations for the structure and combination of the matrix elements determining the 
 $1^-$ transitions, the form of the weak magnetism correction 
that should be used is uncertain.


The finite size correction for the first forbidden $\beta$ decays is a complicated and so far not a satisfactorily
resolved issue. Ideally a simple correction in terms of a formula of $\delta_{FS}$, analogous to that for the allowed
GT decay in the eq. (\ref{eq:finsize}), would be applied to each transition of a given $\Delta J$.
 However, as in the case of the weak magnetism corrections, the finite size 
correction is operator dependent.
Behrens {\it et al.} \cite{behrens} addressed the problem by introducing corrections to the six basic operators, either in terms of
additional radial integrals that have to be evaluated or as tabulated  numerical corrections to the shape factors $C(Z,E)$ \cite{suzuki, zhi}.
 In one application \cite{fang}, the first forbidden decays
of $^{136}$Te and $^{140}$Xe were evaluated both using the shell model and QRPA, and  
the nuclear finite size found to result in a reduction of the neutrino flux above the 1.8 MeV threshold
of 2-5 \% depending on the $E_0$, but to be operator and $\Delta J$ dependent. 

The lack of a comprehensive and/or single
treatment for the nuclear size corrections for forbidden transitions, 
and its detailed dependence on the operators determining the transition,
represents an important source of
uncertainty in the aggregate fission antineutrino spectra.
     
\begin{figure}[htb]
\begin{center}
\includegraphics[width=3.0in]{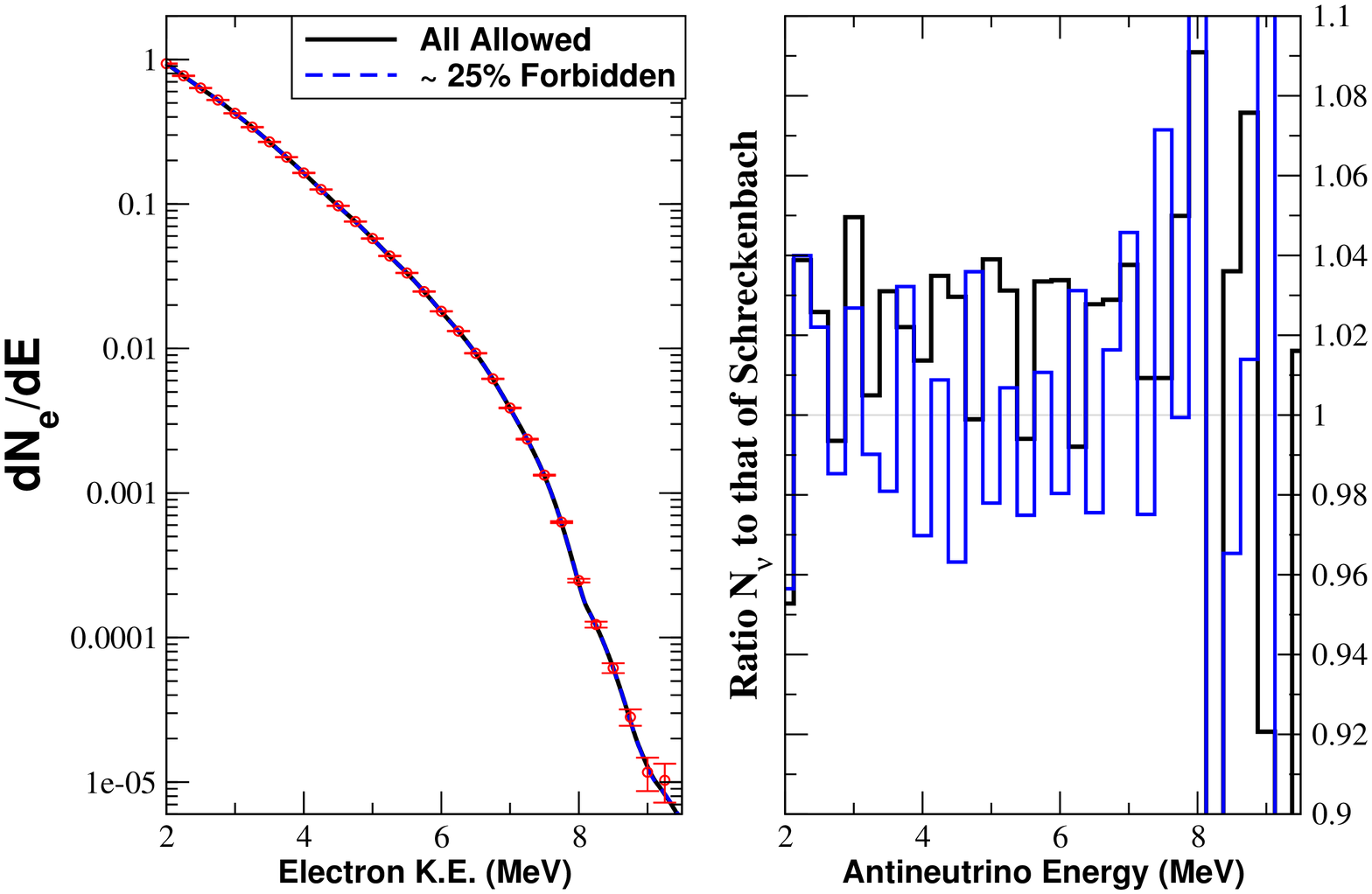}
\caption{ 
The fit to the electron spectrum for $^{235}$U (left) for two
different assumptions on how to treat forbidden transition,
and the ratio of the corresponding antineutrino spectra to
that of Schreckenbach {\it et al.} \cite{Schr2}(right).
 The electron spectrum are fit assuming (a) all allowed
GT branches and (b) 25\% forbidden transitions, and both
$\delta_{FS}$ and $\delta_{WM}$ were included. When
folded over the neutrino detection cross section the case for all
allowed (25\% forbidden) transitions results in a 2.5\% (0.06\%)
increase in the number of detectable antineutrinos. Figure taken from \cite{hayes}.}
\label{forbidden}
\end{center}
\end{figure}

The effect of the forbidden transition operator dependence 
on the deduced antineutrino spectrum using the conversion method has been 
examined in \cite{hayes}. The measured \cite{Schr2} aggregate electron fission spectrum for $^{235}$U was fit
assuming either all allowed transitions or various combinations of the allowed and forbidden operators listed 
in Table \ref{six}.
Excellent fits to the electron spectrum were found in all cases, 
indicating that the electron spectrum cannot distinguish between these scenarios.
 However,
the different treatments of the forbidden transitions lead
to different antineutrino spectra, both in shape and magnitude
at about the 4\% level. 
Two examples, taken from \cite{hayes},  are shown
in Fig.\ref{forbidden}, where in one case all
transitions are assumed to be allowed, while in the second
case the best fit results from about 25\% forbidden
decays. 
For the assumption of all allowed transitions, 
 a systematic increase in the number of antineutrinos
relative to Schreckenbach {\it et al.} \cite{Schr2} of about 2.5\% was seen, 
while in the case
that forbidden transitions were included no increase relative
to that reference is found. 

\section{DETECTOR REACTION $\bar{\nu}_e + p \rightarrow e^+ + n$}


In essentially all experimental studies of reactor neutrino oscillations the 
$\bar{\nu}_e$ capture on protons is the detector reaction
of choice, due to its relatively large cross section and the extremely convenient 
correlated signature of the positron emission followed
by the delayed and spatially correlated neutron capture. Here we briefly review 
the corresponding cross section formulae.
To zeroth order in $1/M$ the cross section is simply related to the neutron decay lifetime
\begin{equation}
\sigma^{(0)}_{tot} = \frac{ 2 \pi^2/m_e^2}{f^R \tau_n} E_e^{(0)} p_e^{(0)} ~
\approx 9.52 \times 10^{-44} 
\left( \frac{E_e^{(0)} p_e^{(0)}}{ {\rm MeV}^2} \right) {\rm cm}^2 ~,
\end{equation}
where $f^R$ = 1.7152 is the neutron decay phase space factor that includes the Coulomb, 
weak magnetism, recoil and outer
radiative corrections, but not the inner radiative corrections,
and $E_e^{(0)} = E_{\nu} - (M_n - M_p)$. 

However, even for the reactor energies the corrections of the first order in $1/M$ 
should be included \cite{BV}
\begin{equation}
E_e^{(1)} = E_e^{(0)} \left[ 1 - \frac{E_\nu}{M}(1 - v_e^{(0)} \cos \theta )\right] 
- \frac{\Delta^2 - m_e^2}{2M} ~,
\end{equation}
where $\Delta = M_n - M_p$ and
\begin{equation}
\left( \frac{ d \sigma}{d \cos \theta} \right) ^{(1)} = \frac {\sigma_0 }{2} 
\left\{ [ (f^2 + 3g^2) + (f^2 - g^2) v_e^{(1)} \cos \theta ] E_e^{(1)} p_e^{(1)} 
- \frac{\Gamma}{M} E_e^{(0)} p_e^{(0)} \right\} ~.
\end{equation}
Here $f = 1$ and $g=1.27$ are the nucleon form factors at $q^2 = 0$, 
$\sigma_0 = (G_F^2 \cos^2 \theta_C)/\pi (1 + \Delta^R_{inner})$ and
$\Delta^R_{inner} \sim$  0.024 is the inner radiative correction.

The quantity $\Gamma$ is given by a somewhat cumbersome formula
\begin{eqnarray}
\Gamma  & = & 2( f+f_2 ) g [ ( 2 E_e^{(0)} + \Delta)( 1 - v_e^{(0)} \cos \theta) - m_e^2/E_e^{(0)} ] + 
( f^2 + g^2 ) [ \Delta ( 1 +  v_e^{(0)} \cos \theta) + m_e^2/E_e^{(0)} ]  \nonumber \\
& & +  [ ( E_e^{(0)} + \Delta ) (1 - \cos \theta / v_e^{(0)} - \Delta] 
\times [(f^2 + 3 g^2 ) + ( f^2 - g^2) v_e^{(0)} \cos \theta] ~.
\end{eqnarray} 
Here $f_2 = \mu_{anom}$ = 3.706 is the nucleon isovector anomalous magnetic moment.

Various forms of extension to all orders in $1/M$ are given in \cite{BV,SV} as well 
as in the classic review \cite{LS}, where, however the threshold behavior is
not properly included. Radiative corrections of order $\alpha/\pi$ were evaluated, 
e.g. in \cite{KMV}. Their convenient numerical form given in
\cite{SV} is
\begin{equation}
d\sigma(E_{\nu} E_e) \rightarrow  d\sigma(E_{\nu} E_e) 
\left[ 1 +  \frac{\alpha}{\pi} \left( 6.00 + \frac{3}{2} \log \frac{M_p}{2 E_e} + 
1.2 \left(\frac{m_e}{E_e} \right)^{1.5} \right) \right] ~.
\end{equation}

\section{THE SHOULDER OR SO-CALLED ``BUMP" IN REACTOR ANTINEUTRINO SPECTRA}

\begin{figure}[htb]
\begin{center}
\includegraphics[width=2.3in]{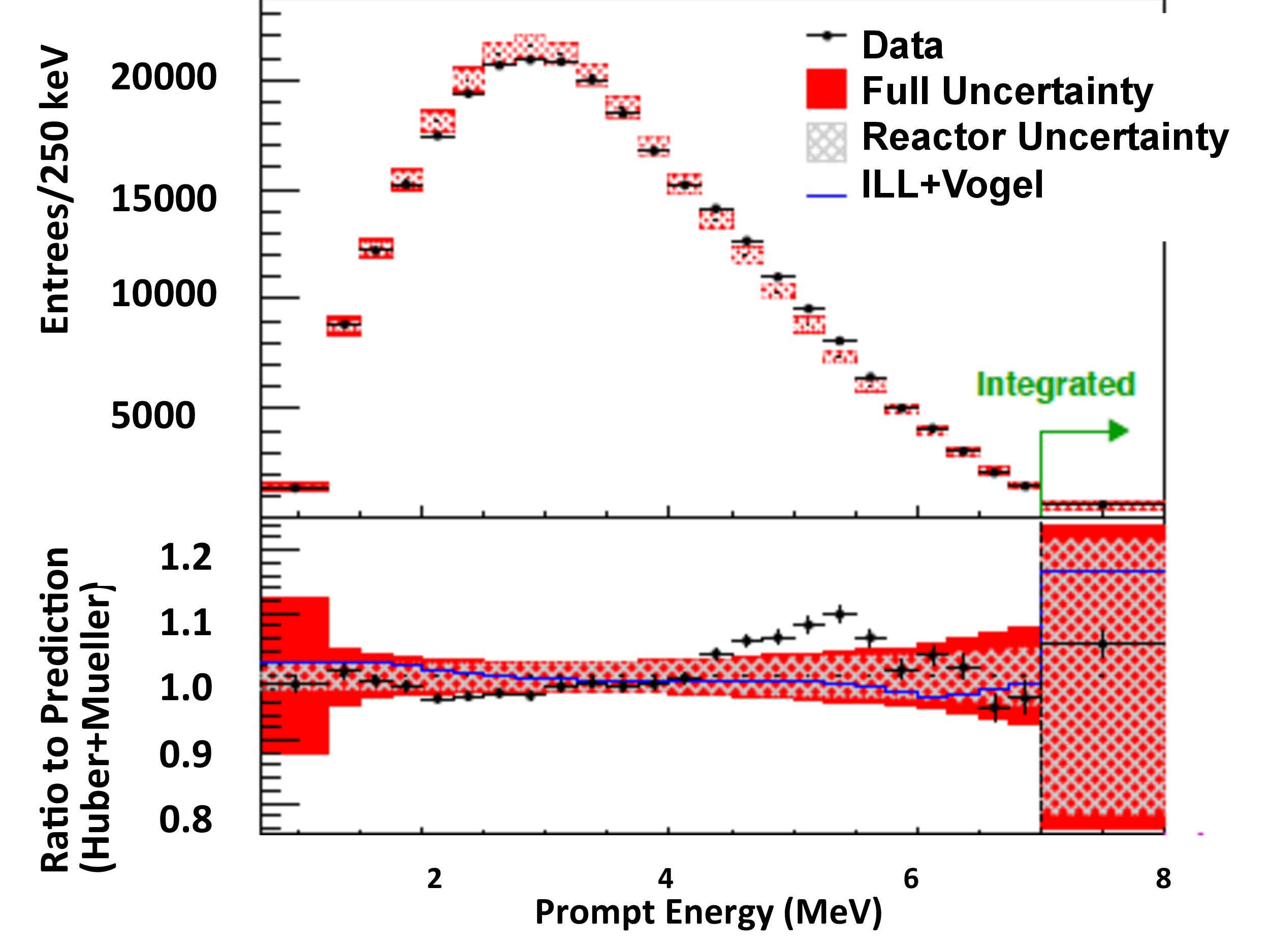}
\includegraphics[width=2.5in]{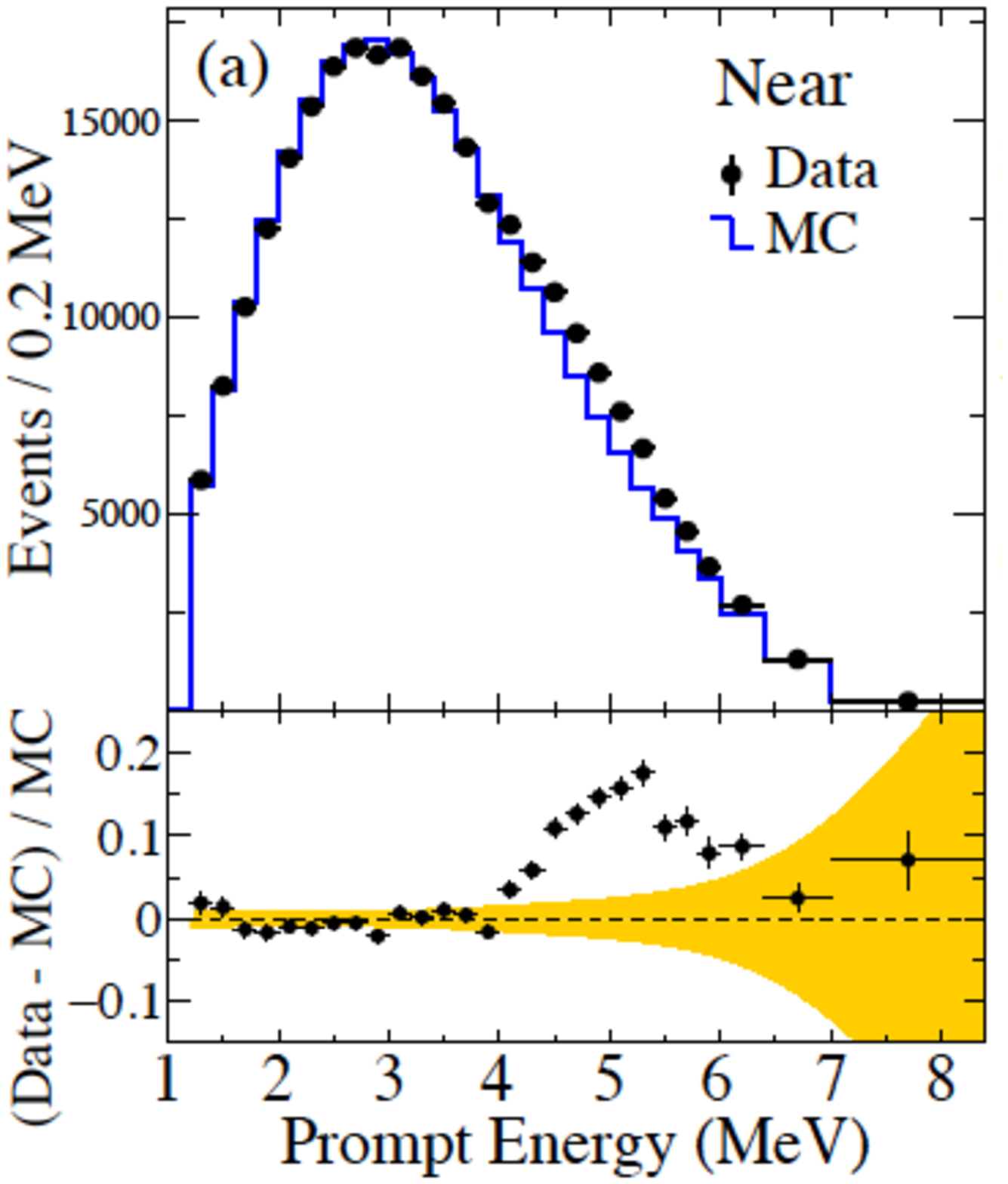}
\caption{(Top) The shoulder (bump) observed in the near detector at Daya Bay \cite{dayabaybump}, arising from
the ratio of the observed spectrum to the predicted.
The blue curve is the ILL prediction \cite{Schr2,Schr3} for $^{235}$U and $^{239,241}$Pu plus
Vogel {\it et al.} \cite{vogel81} for $^{238}$U.
(Bottom) The shoulder observed \cite{seo} in the near detector at RENO.
The predictions are from the Huber-Mueller model \cite{huber, mueller} model, normalized to
the same number of events.
}
\label{fig:bump}
\end{center}
\end{figure}

All three recent large reactor experiments, Daya Bay, RENO, and Double Chooz \cite{dayabaybump,seo,crespo}, observed
a feature (or shoulder) in the 
experimental spectrum at 4-6 MeV of the prompt positron energy, $E_{prompt}\approx E_\nu+(M_p-M_n-M_e)+2M_e$,
relative to the predicted theoretical evaluation in the Refs. \cite{mueller,huber}. The shoulder
 has not been observed in the measured fission electron beta-decay spectra
\cite{Schr2,Schr3,Haag1,Haag}. It was also not observed in the previous test of
the reactor spectrum shape \cite{bugey3}. 
 An example of the data, from the Daya Bay  and RENO experiments,
is shown in Fig. \ref{fig:bump}. In case of Daya Bay, the measured spectrum deviates from the predictions 
by more than 2$\sigma$ over the full energy range and by 4$\sigma$ in the range 4-6 MeV.
The other two experiments (RENO and Double Chooz) report  similar data and similar significance. 
The spectral shape of the shoulder cannot be produced by the
standard $L/E_{\nu}$ neutrino oscillations dependence, independent
of the possible existence of sterile neutrinos. In addition, it is too high in energy
to be produced by antineutrinos emitted from neutron interactions with structural material in the reactor \cite{hayes1} or from the spent fuel. Its origin must be caused by the
reactor fuel $\bar{\nu}_e$ emission.

In the context of the present review several questions need to be considered. 
What is the origin of this ``bump"? Why was it not observed in the electron
spectrum? Does its existence question expected/predicted reactor spectra in general?

The shoulder could have its 
origin in several effects that are not included, or not included accurately, in the reactor spectrum
predictions \cite{mueller,huber}. Many of the important decays are forbidden, so that 
their shape factors and sub-dominant corrections might be different than assumed.  Alternatively,
the contribution
of $^{238}$U, that is only weakly constrained by the observed electron spectrum 
might not be accurate.  The harder neutron spectrum in
power reactors may lead to
different fission fragment distributions than in the very thermal ILL reactor used for the electron fission spectra measurements
\cite{Schr2,Schr3,Schr4}.
Finally, the measured electron spectra themselves \cite{Schr2,Schr3,Schr4},
which represent the basis for the antineutrino evaluations \cite{mueller,huber}, might
be incorrect.  

The reactor $\bar{\nu}_e$ spectra are composed of $\beta$ decays of hundreds of individual 
fission fragments, with $\sim$ 6000 individual decay branches. However, at the relevant 
energies, 4-6 MeV of prompt energy, corresponding to $\sim$4.8-6.8 MeV  $\bar{\nu}_e$
energy, relatively few ($\sim$ 10-15) transitions determine 40-50\% of the
total spectrum, refs. \cite{sonzogni,sonzogni1,dwyer}, and they are mostly forbidden transitions. 
The other more numerous decays that determine the remainder of the spectrum in the bump energy window each contributes less than 2\%. 
It is, therefore,
possible that the conversion from
the electron to the high-energy component of antineutrino spectrum involved an inaccuracy that resulted in a shoulder.

Several possible origins of the bump have been identified and investigated by different authors \cite{dwyer, hayes1},
but it was generally concluded that, without further experimental investigation,
it is impossible to determine which, if any or several, of the explanations are correct. 
However, several comments are in order, and we summarize the situation here.
 
Dwyer and Langford \cite{dwyer} used the {\it ab initio} summation method to construct the 
electron and antineutrino spectra from  the ENDF/B-VII.1 fission yield and decay libraries,
assuming allowed shapes and including the corrections discussed above. They observed that the ENDF/B-VII.1
library predicted  a shoulder
or ``bump" very similar to that observed \cite{dayabaybump,seo,crespo}. 
In addition, they showed that a corresponding bump was predicted relative to the 
original measured \cite{Schr2,Schr3,Schr4} aggregate fission electron spectra.
This  explanation would, therefore, suggest that the measured electron spectra  are
incomplete, i.e. the shoulder was missed somehow in the measurements. 

\begin{figure}
\includegraphics [width=2.5in] {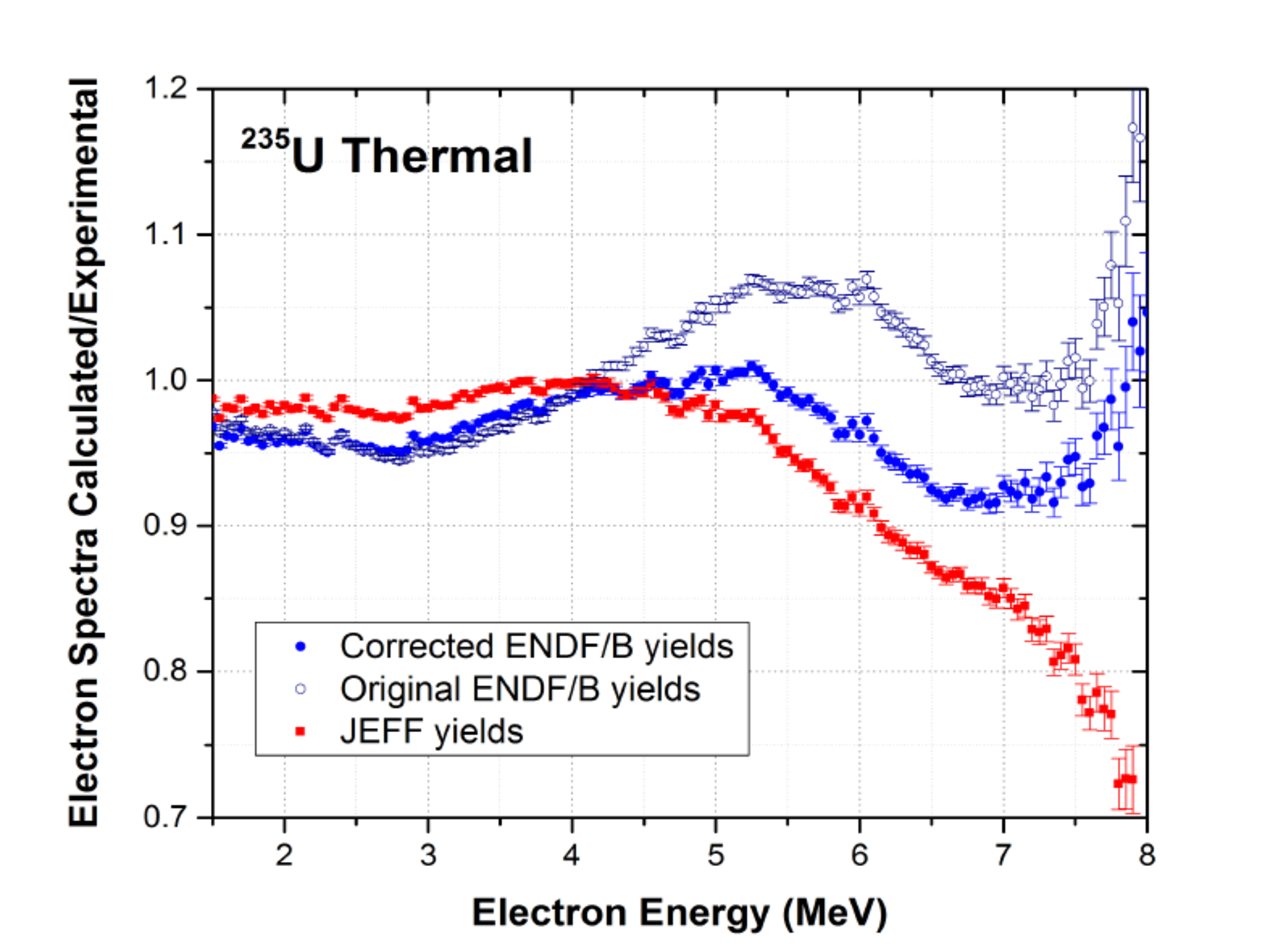}
\caption{The ratio of the database predictions to the measured \cite{Schr2,Schr3} electron spectrum for $^{235}$U.
Neither JEFF nor the corrected ENDF/B
database predicts a bump relative to the measured electron spectrum. 
Figure reproduced with the permission of A. Sonzogni {\it et al.}}
\label{fig:sonzogni}
\end{figure}

In the {\it ab initio} summation method the necessary input are the fission yields, and 
two standard fission-yields libraries, JEFF-3.1.1 and ENDF/B-VII.1
differ \cite{hayes1} significantly in the
predicted yields of several nuclei dominating the shoulder region.
In particular, 
the JEFF-3.1.1 library fission yields  does not predict \cite{hayes1} a ``bump" for the Daya Bay or RENO experiments, and agrees
reasonably well with the measured electron spectra.
A recent critical review \cite{sonzogni1} of the ENDF/B-VII.1  yields for $^{235}$U uncovered erroneous yields for $^{86}$Ge
and all of its daughters, and showed that this 
error was generating excess of strength in both the predicted electron and antineutrino spectra
 at 5-7 MeV. When this problem was corrected, along with other less critical updates to the library,
the predictions of the two databases are considerably closer, 
and agree within 6\% at all energies. 
Most significantly, neither database (corrected ENDF or JEFF)
now predict a bump relative to the measured $^{235}$U aggregate electron fission spectrum, Fig. \ref{fig:sonzogni}.
Thus, at present, there is no evidence that the original measurements of the electron spectra are the origin of the bump.

An alternate source of the bump might lie with 
the conversion of the measured electron spectra to antineutrino spectra, which would point 
to the shoulder being
produced by the approximations made for the corrections and/or for the forbidden transitions. 
However, this latter  possibility can be mostly ruled out, 
(see ref. \cite{hayes1}), 
by examining the expected change in the bump region of spectrum for different treatments of the forbidden transitions.
For example, three fission fragments that dominate in the bump region, $^{92}$Rb, $^{96}$Y and $^{142}$Cs, 
all involve $0^-\rightarrow0^+$ decays and, thus, 
have no weak magnetism correction. Since the weak magnetism correction is opposite in sign to the finite size correction, a proper treatment of weak magnetism is important. 
There are two $\Delta J=0^-$ operator, one GT and one proportional to the axial charge $\rho_A$, each with different shape factors, $C(E_e)$. Thus, there is considerable uncertainty as to how to treat these transition.
To test this no weak magnetism and no shape correction was applied in one case to these three transitions and.
In the second case no weak magnetism and the shape correction for a purely GT 0$^-$.
The first treatment leads to a small increase in the antineutrino spectrum above about  4 MeV, which is a maximum of 1\% at 8 MeV, while the second leads to a suppression in the energy region of interest. Thus, it was concluded in ref. \cite{hayes1} that 
 a proper treatment of weak magnetism for forbidden transitions
cannot account for a significant fraction of the shoulder.

\begin{figure}
\includegraphics [width=2.5in] {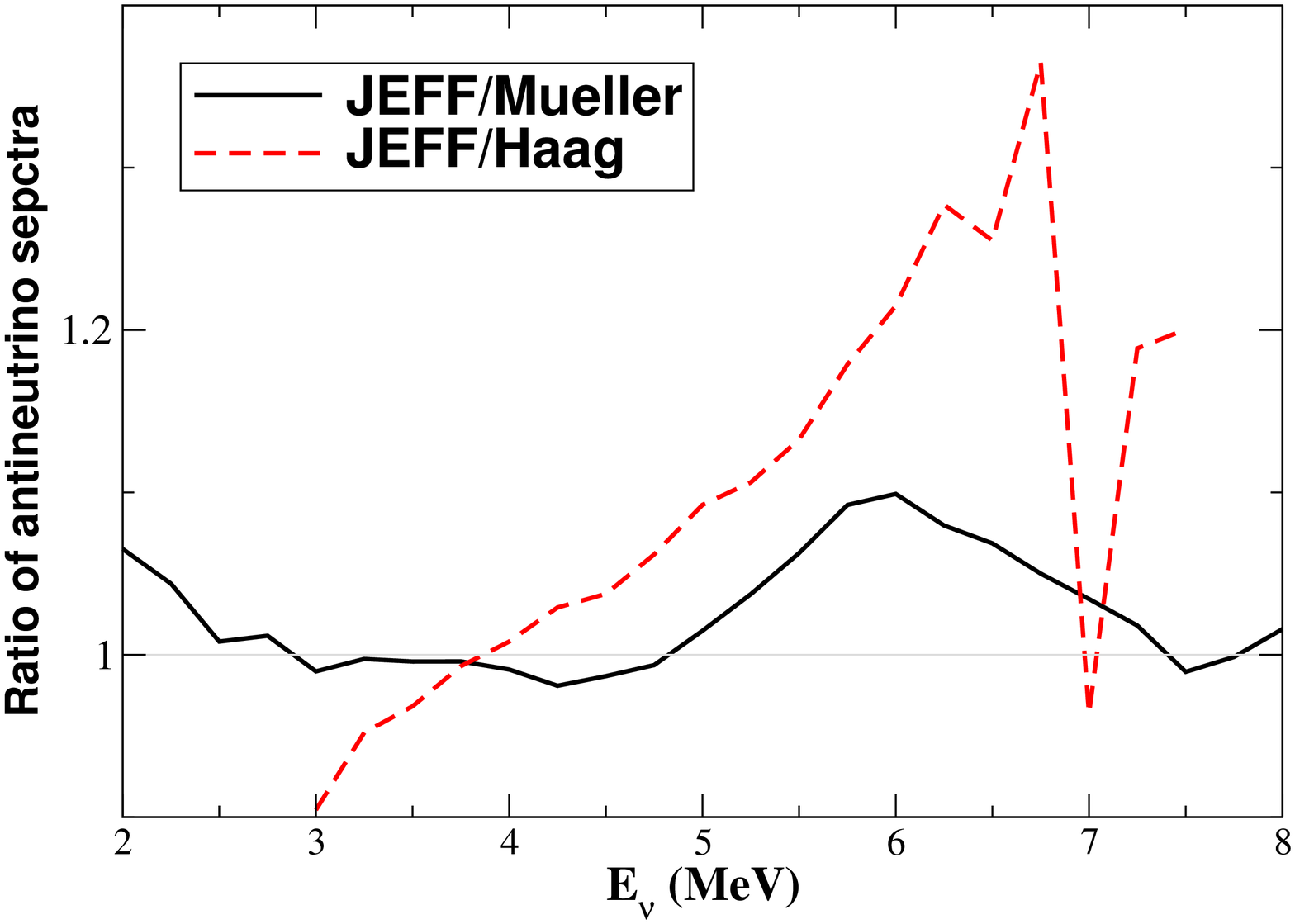}
\caption{The ratio of the JEFF-3.1.1. antineutrino spectrum for $^{238}$U to that of 
Mueller \cite{mueller} and Haag \cite{Haag}.
As can be seen,JEFF predicts a  bump relative to the predictions of both Mueller and Haag, with the latter bump being the larger. Double Chooz use the Haag $^{238}$U antineutrino spectrum, and JEFF-3.1.1 predicts a bump relative to Double Chooz, see ref. \cite{hayes1}}.
\label{238}
\end{figure}

At present, the two most likely sources of the bump seem to be $^{238}$U or the hardness of the neutron spectrum.
In the case of $^{238}$U, there are a few observations worth commenting on.
First, $^{238}$U represents about 12\% of the total fissions at RENO,
compared to 7.6\% and 8.7\% at Daya Bay and Double Chooz, respectively,
and the bump seen at RENO is larger than in the other two experiments.
Also, the $^{238}$U spectrum is considerably harder in energy than that of the other actinides, 
which results in $^{238}$U contributing about 
24\% (15\%) of the spectrum in the bump region for RENO (Daya Bay).
Second, {\it both} the ENDF/B-VII.1 and JEFF-3.1.1 libraries predict a bump
relative to the $^{238}$U antineutrino spectrum of Mueller \cite{mueller} and of the recent 
measurement of 
Haag \cite{Haag}, as shown in Fig. \ref{238}.
Thus, without
experiments designed to isolate the contributions from
each actinide to the shoulder, $^{238}$U cannot be ruled out as a significant source of the bump.

 Finally, the effect of the hardness of reactor neutron spectrum on the
antineutrino spectrum has never been tested directly.
The databases generally predict this to be a small effect.
Nonetheless, the PWR reactors used by Daya
Bay, RENO and Double Chooz are harder in energy than
the thermal spectrum of the ILL reactor, and involve
considerably larger epithermal components. 
In the epithermal energy region the fission yield distributions can be resonance dependent \cite{hambsch}.
If epithermal fission is a significant source of the bump, one might expect it
to be most pronounced in $^{239}$Pu, since there  the first epithermal fission 
resonance at $E_n$=0.32 eV is quite isolated and strong, and it 
can account for as much as 25\% of the total plutonium fission
in some pressurized water reactors. 
Thus, any experimental 
tests of the variation of the yields of the dominant fission fragments
with neutron energy would be very valuable in addressing this issue.

The existence of the ``bump" has little effect \cite{ciara} on the extraction of the neutrino oscillation
parameters from the reactor experiments. In addition, it could be entirely
uncorrelated with the ``reactor anomaly".
However, it raises the very serious question of how well the antineutrino spectra are known,
and suggests that estimated uncertainties at the 1-2\% are too optimistic.

\section{THE REACTOR ANOMALY AND NEW EXPERIMENTS}

\begin{figure}[htb]
\begin{center}
\includegraphics[width=8.0in]{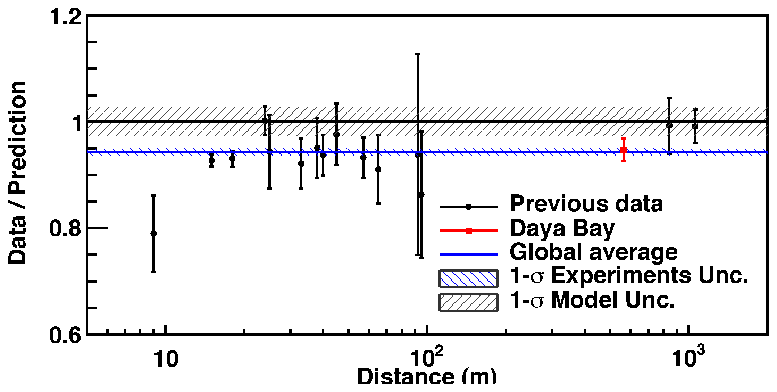}
\caption{The measured reactor rate, normalized to the prediction of \cite{mueller,huber} plotted
as a function of the distance from the reactor core. The rate is corrected for the 3-flavor
neutrino oscillation at each baseline. The data at the same baseline are combined for
clarity.  The Daya-Bay measurements are shown at the flux-weighted baseline  (573 m)
of the two near halls. Reproduced with permission from \cite{dayabaybump}.}
\label{fig:anomaly}
\end{center}
\end{figure}

The total yield of $\bar{\nu}_e$ capture on protons
 measured in all past reactor experiments is lower than the predictions of  Refs. \cite{mueller,huber},
a finding which is generally referred to as the ``reactor neutrino anomaly".
The present status is illustrated in Fig. \ref{fig:anomaly} where the older data are shown, 
together with the more recent high statistics  result from the Day-Bay experiment
\cite{dayabaybump}. For the Daya-Bay experiment alone the ratio of the measured to expected yield
is  0.946$\pm$0.022.
A global fit, that includes all past measurements, results in the ratio
$R = 0.942 \pm 0.009 {\rm (exp)} \pm 0.025 {\rm (model)}$, when corrected for the
known neutrino three-flavor oscillations. The average value of $R$
is well determined, and the experimental
uncertainty is 
substantially smaller than the model uncertainty assumed in \cite{mueller,huber}.  
Taking the quoted uncertainty in the model predictions at face value,
the global value of $R$ suggests reactor $\bar{\nu}_e$ disappearance as close as $L <$ 10 m. Such short baseline disappearance cannot be 
accommodated within the standard $3-\nu$ mixing model, and hence the phrase 'reactor anomaly'.

The reactor anomaly 
is one of several experimental results that contradicts the 
standard three-flavor neutrino oscillation 
paradigm. 
Other experiments, including LSND \cite{lsnd}, MiniBooNE \cite{miniboone}, SAGE \cite{sage} and GALLEX \cite{gallex}, 
indicate non-standard $\nu_e$ disappearance or $\nu_{\mu} \rightarrow \nu_e$ 
appearance, albeit with the statistical significance of only $\sim 3 \sigma$.
The common feature of such anomalies is that the parameter $L {\rm (m)}/E_{\nu} {\rm (MeV)}$ is of the
order of unity, and  
this similarity in $L/E_{\nu}$ has led to a proposed explanation involving one or more additional neutrino types, i.e.,
sterile neutrinos. Clearly, this is an issue of fundamental importance,
potentially  a source of the long sought after ``physics beyond the standard model". 
  
It is beyond the scope of the present review
to describe all of the experiments that comprise the current neutrino anomalies and 
their  interpretation. 
However, detailed discussions, including a description of the experiments and their analyses, are provided in  Refs. \cite{kopp,giunti}.
The possible existence of sterile neutrinos
has to be confirmed or ruled out  with new experiments, and a  comprehensive discussion
of the issue  is provided in the white paper \cite{abazajian}. 
Here we concentrate on aspects of the
problem related to reactor neutrinos, where the experimental data are quite firm, but the expectations
depend on an assumed reactor spectrum,  involving uncertainties that are difficult to determine reliably.
These uncertainties and the associated complications are discussed in a separate section 
of this article.
 
Quite generally,
$|\Delta m^2_{i,j} {\rm (eV^2)}| \cong 2.48 E_{\nu} {\rm (MeV)}/L_{osc}{\rm (m)}$.
Thus, $L/E_{\nu} \sim 1$ implies $ \sim \Delta m^2_{i,j}$ of a few eV$^2$ and the corresponding additional
neutrino (or neutrinos), if they exist, must have the mass $m_{sterile} \sim 1$ eV, assuming that
$m_1,m_2,m_3 \ll m_{sterile}$. The analysis in \cite{giunti}
restricts $\Delta m^2_{4,1}$ to the interval 0.8 - 2.2 eV$^2$. The extra sterile neutrinos would have
experimental consequences and their existence would be observable only if they mix with the active neutrinos. 
If sterile neutrinos are indeed responsible for current anomalies in neutrino physics,
 the corresponding mixing 
angle is $\sin^2_{new}2\theta \sim 0.1$, with a range of about 0.06 - 0.22.

 To test the short baseline oscillation sterile neutrino hypothesis it is necessary 
 to perform experiments  sensitive not only to the total flux, but more importantly
 to the oscillation pattern in $E_{\nu}$ or $L$. 
 For reactors this translates into the distance between the neutrino source and the detector
 being between $\sim$1-15 meters; at larger distances 
 the oscillatory behavior would be washed out. 
 The statistical and systematic uncertainties of new experiments
 have to be at the level of few percent for a definitive result.
 A number of reactor experiments that fulfill these criteria have been proposed. They are at different
 stages of planing, funding, and construction.
The same sterile neutrino hypothesis can be also tested 
 with strong radioactive sources at similar distances or with accelerator neutrinos at proportionally larger distances,
 and there are also a number of proposed experiments of this type. 
 
 The new reactor experiments, with planned high statistics and low systematic uncertainties,  would also provide
 very significant and valuable tests of the reactor antineutrino spectrum in general. 
Some of the experiments will use research reactors based on the highly
 enriched $^{235}$U fuel, and have the additional advantage of a compact-sized reactor core. 
 Thus, the $^{235}$U spectrum can be isolated and well-determined, and if statistics are good enough, 
 the ``bump" feature explored. 
Experiments are planned at highly thermal neutron flux reactors and at higher neutron temperature reactors, 
and comparisons between these
will be important in shedding light on the role of epithermal neutrons in determining the antineutrino spectrum.  
When combined with the results from power reactors it should be possible to isolate the
 spectra of the other reactor fuels to much  better accuracy that is currently possible. 
The results of such tests would also benefit the
 Applied Antineutrino Physics \cite{bowden} aimed at monitoring reactor operations and fuel content.

 Multiple very short-baseline reactor (VSBR) experiments are described together in Refs. \cite{lasserre,nature-reactor}. They have been proposed globally; in the US (PROSPECT, NuLat), Belgium(SoLid), France (NUCIFER, STEREO),
 Russia (DANSS, NEUTRINO-4, POSEIDON), and Korea (HANARO). 
 Generally, since the VSBR experiments are at shallow depth, and necessarily near the reactor cores,
 control of the  backgrounds is a challenging task. 
The detectors are usually Gd loaded or $^6$Li loaded liquids.
 While the Gd-liquid scintillator technology is mature, the $^6$Li has the advantage that the neutron capture
 produces an $\alpha$ particle and triton. This provides a good localization of the delayed signal and additional
 pulse-shape discrimination. 
 Within few years of data taking, the parameter region suggested by the anomalies should be well-covered
 and the experiments should be in a position to make definitive statements on the possible existence of
 light sterile neutrinos.

\section{UNCERTAINTIES IN THE ANTINEUTRINO SPECTRA}

For many neutrino oscillation analyses the uncertainties 
in the expected antineutrino spectra are as important as the spectra themselves.
As discussed above, the expected spectra can be derived by two main methods,
by summing all of the individual transitions that make up the spectrum using the nuclear databases as input, 
or by converting a measured aggregate fission electron spectrum to an antineutrino spectrum.
The issues determining the uncertainties are somewhat different in the two case.

The summation method requires knowledge of both the decay spectra and fission yields for all of the fragments determining the spectra, and both inputs involve uncertainties.
For the decay of individual nuclei, the databases are incomplete because about 5\% of the nuclei are 
sufficiently far from the line of stability that no measurements of the spectra are available.
Thus, summation methods must rely on some modeling to account for the missing spectra.
A model for the  spectrum of each missing nucleus is provided by ENDF
in terms of a continuous spectrum, which is derived from nuclear structure calculations.
The model \cite{shannon, kawano} is an extension
of the Finite-Range Droplet Model plus Quasi-particle Random Phase Approximation,
tuned to account for the so-called pandemonium effect \cite{Hardy}  (viz., a very large number of low-energy beta decays to the many high-lying excited states of the daughter) as well as forbidden transitions,
and is supplemented by the nuclear structure library ENSDF \cite{ENSDF} where appropriate.

For many of the nuclei with decay measurements, 
the decay schemes are uncertain and the spin and parity $\Delta J^\pi$ involved in the transition is unknown for many branches. 
In addition, since about 30\% of the transitions are forbidden, the finite size $\delta_{FS}$ and weak magnetism $\delta_{WM}$ corrections and  the shape factors $C(E,Z)$ applied to these transitions 
must be assigned large uncertainties.
For all forbidden transitions the finite size has been approximated to-date.

The radius determining the finite size correction involves both the weak transition density $\rho_W$ and the charge density $\rho_{ch}$.
Under the assumption that $\rho_W=\rho_{ch}$, 
we recover eq. (\ref{holFS}).
However, the radius describing 
the finite size corrections is nuclear structure dependent.
 Several density or radius approximations 
have been made in the literature \cite{dicus, biottino, armstrong}, and these differ from one another by about  50\%.
Thus, we place a 50\% uncertainty on the allowed finite size correction.
For forbidden transitions, a general analytic expression for the  form of the finite size corrections has not been derived. A treatment to leading order in $q^2$ has been derived in \cite{behrens}, but it involves calculating several nuclear structure dependent matrix elements. Thus, we place a 100\% uncertainty on the finite size correction to forbidden transitions.

The weak magnetism correction also involves some uncertainty. There are two-body meson-exchange current corrections that are
nuclear structure dependent.
These are typically of the order of 10-20\% corrections. In addition, as the weak magnetism correction involves 
matrix elements of both $\sigma$ and $\ell$, and, in the absence of detailed nuclear structure calculations, some assumption must be made about $\langle \ell \rangle$. For the allowed and uniquely forbidden weak magnetism corrections we assign a correction of 20\%. For the non-unique forbidden $1^-$ transitions we suggest an uncertainty of 25\% to take account of the fact
that $\delta_{WM}$ depends on the operator.
In the case of $\Delta J= 0^-, 1^-$, more than one shape factor can contribute and the combination is nuclear structure dependent.  Based on comparisons of the total antineutrino spectra computed using different choices of these $C(E,Z)$, we estimate the uncertainty to be about 30\%. 

The database fission yields are also
uncertain for many important nuclei, and differences \cite{hayes1} between the JEFF-3.1.1 and ENDF/B-VII.1 evaluated yields of these nuclei were found to be as much as 20-50\%.
A more recent review and revision 
of  the important fission yield in the ENDF/B-VII.1 database by Sonzogni {\it et al.} \cite{sonzogni1}
 has brought antineutrino
spectrum predicted for $^{235}$U by JEFF and ENDF databases to within better than 6\% of each other over most all of
the energy
range relevant to reactor neutrino experiments.
The two databases do deviate significantly between $E_\nu=$7-8 MeV, but this energy window represent only a small fraction of the observed total.
Up to about 4.5 MeV, ENDF/B is lower than JEFF, but above this energy JEFF drops steadily, becoming more than a factor of 1.5 lower than ENDF/B.
Both databases are lower than the measured antineutrino spectrum for $^{235}$U over the energy range $E_e=1.5-7.5$ MeV, although this may reflect the need to further correct the databases for the so-called   pandemonium effect.
While it is difficult to estimate the uncertainty in the database fission yields, we tentatively place the uncertainty arising from their contribution to the summation method at $\sim 10\%$, motivated 
in part by the good comparisons to decay-heat \cite{decayheat} and  the work of \cite{sonzogni1}.

\begin{table}
\caption{ The estimated uncertainties for the ingredients that make up the aggregate antineutrino spectra when the summation method is used. These estimates are subjective and are bases on the the educated guess of the authors. They do not represent
statistical variances.}
\begin{tabular}{c|c|c|c}
Quantity&type&$\Delta J^\pi$& uncertainty\\\hline 
Unknown branching and $J^\pi$&allowed and forbidden&all&50\%\\
Finite size corr.&allowed&$1^+$&50\%\\
Finite size corr.&forbidden&$0^-,1^-,2^-$&100\%\\
Weak magnetism&allowed&$1^+$&20\%\\
Weak magnetism&forbidden&$0^-$&0\\
Weak magnetism&forbidden&$2^-$&20\%\\
Weak magnetism&forbidden&$1^-$&25\%\\
Shape factor&allowed&$1^+$&0\\
Shape factor&forbidden&$2^-$&0\\
Shape factor&forbidden&$0^-,1-$&30\%\\
Fission yields&allowed and forbidden&all&10\%\\
Missing spectra&allowed and forbidden&all&50\%\\\hline
\label{uncert}
\end{tabular}
\end{table}

Though many of the uncertainties in Table \ref{uncert} also apply when converting a measured electron spectrum to an antineutrino spectrum, the situation is somewhat different. This is because the fit must reproduce the electron spectrum, even if the finite size correction is changed by, say, 50\%. 
In ref. \cite{hayes} different assumptions were made about which weak magnetism and shape factors should be applied to the non-uniquely forbidden component of the spectrum and with fits to the electron spectrum of equal statistical accuracy, the antineutrino spectrum was found to vary by 4\%. In the latter work, only expressions listed in  Table \ref{six} were used for
$\delta_{WM}$ and $C(E,Z)$, i.e., no uncertainty was included. In addition, the finite size correction was kept as in eq. (\ref{holFS}). 
To determine the full effect of the uncertainties that apply to a conversion from a measured electron spectrum to an antineutrino spectrum,  
listed in Table \ref{uncert} (not including the last two listings), requires a detailed multi-parameter sensitivity 
study. In the absence of such a study, we tentatively place a 5\% uncertainty on
the conversion method. 

\section{SUMMARY AND FUTURE DIRECTIONS}

Nuclear reactor neutrino experiments have played a central role in neutrino physics since the 1950s.
Despite the complexity of the spectra of antineutrinos emitted from reactor, 
namely, that they result from
thousands of beta-decay branches of unstable fission fragments, 
the spectra were determined reasonably accurately already in the 1980s.
However, today's neutrino oscillation studies have reached a precision such that there is a need to
know the spectra to much higher accuracy, i.e., to considerably better than 5\%. 
For example, 
the reactor neutrino anomaly which suggests the existence of a $\sim 1$ eV sterile neutrino,
represents a 6\% discrepancy between expected and observed number of detected antineutrinos 
in all short baseline experiments. The total signal rate is experimentally 
determined to better than 1\% accuracy.  The statistical significance of the discrepancy, however, crucially
depends on the uncertainty in the expected spectra.

In this review we have attempted to summarize the experiments and models that have constituted 
the ``expected" spectra and how these have evolved over the years.
Determining the uncertainties in the expected spectra is quite difficult and there are many issues involved.
In general, conversion of measured aggregate electron fission spectra provide more accurate determinations of the
antineutrino spectra than do predictions from the databases. 
The database calculations do, however, provide a very important means of estimating the relative importance
of theoretical corrections to the spectra and their uncertainties.
 In examining these and the underlying theories used to derive the corrections, we estimate that the
uncertainty in antineutrino spectra derived by the conversion method are about 5\%. 
The uncertainties in the summation method are considerably worse and at least in the upper part of the antineutrino spectrum are  probably up to the 20\% level.
We emphasize that these are 
our subjective estimates. They are based on educated guesses and they  do not represent statistical variances.

Improving on the theoretical inputs to the spectra will be challenging. Thus, there is a clear need for new experiments.
Ideally, these should involve more than one reactor design and fuel enrichment, because the remaining issues will require
an better understanding of the role of the hardness of the reactor neutron spectrum and of  the four individual
actinides that make up total spectra.  
For the bump energy region, better measurements  of the $^{238}$U spectrum would be particularly valuable.   

\section*{ACKNOWLEDGMENTS}
Anna Hayes thanks the LDRD program at Los Alamos National Laboratory for partial support for this work.
Petr Vogel thanks the Physics Department at the California Institute of Technology for partial support for this work.
They both thank Jim Friar, Gerry Garvey, Alejandro Sonzogni, Libby McCutchen, Jerry Jungman, and members of the 
Daya Bay, Double Chooz, and RENO experimental teams for valuable discussions.
%

\end{document}